\documentclass[fleqn,twoside,twocolumn,nofootinbib,showkeys]{revtex4} % Specifies the document class %,unsortedaddress
\usepackage[nocpr]{ujp} % \usepackage[cyr]{ujp} for cyrillic, \usepackage[web]{ujp} for web
\begin{document}
\title[Structure Functions of Many-Boson System]%колонтитул
{STRUCTURE FUNCTIONS OF MANY-BOSON\\ SYSTEM WITH REGARD FOR
DIRECT THREE- \\
AND FOUR-PARTICLE CORRELATIONS}%
\author{I.O. Vakarchuk}%1 автор
\affiliation{Ivan Franko National University of Lviv}%
\address{12, Dragomanov Str., Lviv 79005, Ukraine}
\email{chair@franko.lviv.ua}
\author{O.I.~Hryhorchak}%1 автор
\affiliation{Ivan Franko National University of Lviv}%
\address{12, Dragomanov Str., Lviv 79005, Ukraine}
\email{HrOrest@gmail.com}

\udk{538.941}  \razd{\secvi}

\autorcol{I.O.\hspace*{0.7mm}Vakarchuk,
O.I.\hspace*{0.7mm}Hryhorchak}

\setcounter{page}{1115}%

\begin{abstract}
On the basis of the expression for the density matrix of interacting
Bose particles in the coordinate representation with regard for the
direct three- and four-particle correlations [I.O.\,\,Vakarchuk and
O.I.\,\,Hryhorchak, J.\,\,Phys.\,\,Stud.\,\,\textbf{3}, 3005
(2009)], the two-, three-, and four-particle structure factors of
liquid $^{4}$He in a wide temperature interval were calculated in
the approximation ``one sum over the wave vector''.\,\,In the
low-temperature limit, the expression obtained for the two-particle
structure factor transforms into the well-known one.\,\,In the
high-temperature limit, the expressions for the two-, three-, and
four-particle structure factors are reduced to those for the ideal
Bose gas.\,\,The results obtained can be applied to calculations of
the thermodynamic functions of liquid $^{4}$He and to the
determination of the temperature dependence of the first-sound
velocity in a many-boson system.
\end{abstract}


\maketitle

\section{Introduction}

The researches of structure functions play an important role in
studying the Bose and Fermi systems, because the results obtained
theoretically can be compared directly with experimental
data.\,\,The central role in the structural researches of those
systems belongs to the total scattering cross-section, which is
called the dynamic structure factor.\,\,This parameter makes it
possible to determine both the spatial structure of the substance
and the structure of its energy spectrum \cite{BPS,FeFam}.\,\,With
its help, as well as with the help of its derivatives, a lot of
different systems are studied today, e.g., the Bose gas in a trap
\cite{ZPSS}, liquid $^{4}$He \cite{KrLi} and $^{3}$He \cite{HHKP} in
two dimensions, solid $^{3}$He \cite{SPA}, thin films \cite{KrTy},
Lennard-Jones rarefied gas \cite{MiSc}, superfluid helium
\cite{BTV}, parahydrogen \cite{DBRF}, models with turbulence
\cite{HaJa}, and so forth.

Besides the dynamic structure factor, not less important is its
zeroth moment or the static structure factor, which has been
measured a lot of times in a wide temperature interval.\,\,The
researches were carried out at the saturated vapor pressure within
the neutron \cite{Svensson} and X-ray \cite{Robkoff} diffraction
methods.\,\,Experimental works on the structure factor measurement
were analyzed, e.g., in work \cite{CBA}, where corrections were also
proposed in order to coordinate results of various authors.\,\,The
Monte-Carlo method was applied to study the region around the
structure factor peak in a Bose condensate, and the peak was shown
to locate higher than the theoretical value obtained in the
framework of the low-density approximation \cite{SOKD}.\,\,The
contribution of three-particle correlations to the structure factor
of liquid $^{4}$He was found in work \cite{ChCo}, and a procedure of
calculation of the effective pair potential on the basis of
experimental data obtained for the structure factor and with the
help of the Monte-Carlo simulation scheme was proposed in work
\cite{ALM}.\,\,Among structure functions, we also mention the pair
correlation function, which is one of the key quantities
characterizing the coherent properties of a Bose condensate
\cite{ZTG}.

In this work, we aimed at finding not only the pair structure
factor, but also expressions for the three- and four-particle
structure factors in the approximation ``one sum over the wave
vector''.\,\,This result should help us, in turn, to simplify
calculations of the thermodynamic functions of a Bose system in the
approximation \textquotedblleft two sums over the wave
vector\textquotedblright\ and to facilitate the
solution of the still unsolved task to describe such a system as liquid $^{4}%
$He in a wide temperature interval and, especially, in a vicinity of
the $\lambda$-transition.

The structural properties of liquid $^{4}$He at low temperatures
have been discussed for a long time in the framework of the
collective variable approach
\cite{VakUhn79-80,VHU79,Vak85-90,VH1,Hlushak}.\,\,However, a
theoretical calculation of the pair structure factor in a wide
temperature interval has been carried out only recently in work
\cite{VPR2007}, by taking advantage of the averaging with the
density matrix of interacting Bose particles.\,\,Later on, the
structure functions in a wide temperature interval were also
described in other works \cite{VP2008-09}.\,\,However, the authors
of the cited works calculated the average values in the framework of
the density matrix approach and in the pair correlation
approximation, so that the pair structure factor was obtained in the
same approximation.\,\,The agreement with experimental data for the
pair structure factor \cite{Svensson,Robkoff,exps1} is good in this
case, but incomplete, because, as is known, the contribution of
many-particle correlations to the observed quantities of a
many-boson system can turn out rather substantial
\cite{Temperly,Krokston,VH1}.

In works \cite{VP2008-09}, the irreducible two-, three-, and
four-particle structure factors, as well as the pair distribution
function, were calculated in a wide temperature interval making
allowance for only the indirect three- and four-particle
correlations.\,\,The obtained theoretical results can be improved by
taking direct correlations into account as well.\,\,However, in this
case, the indicated quantities have to be calculated with the
density matrix containing not only the pair, but also three- and
four-particle direct correlations.\,\,This task is a purpose of this
work.\,\,In our calculations, we will base on the approaches
proposed in our earlier works \cite{VakHryh1,VakHryh2} and the
results obtained there; in particular, these are the expressions for
the density matrix and the partition function for a many-boson
system in a wide temperature interval and the methods of their
calculation in the approximation ``two sums over the wave vector''.

An important feature of this work is a graphic presentation of the
results obtained.\,\,As a rule, numerical calculations are carried
out for this purpose.\,\,The input data at such calculations include
experimental results for the structure factor extrapolated to the
absolute zero temperature.\,\,The general scheme of speculations on
this topic and the corresponding results can be found in work
\cite{VBR}.\,\,Continuing the issue of numerical calculations, it is
worth paying attention to work \cite{Rov-dys}, where the interatomic
interaction potentials were restored on the basis of experimental
data (as was done in work \cite{VBR}), and the thermodynamic and
structural properties of $^{4}$He were \mbox{studied.}%\looseness=1

The numerical calculation of the pair structure factor was carried
out taking the effective mass into account.\,\,The expression for
the latter was given in work \cite{HP2014}.\,\,A necessity of its
introduction was substantiated in work \cite{VP2008-09}.

The expression obtained for the two-particle structure factor
transforms into an already found one for the low-temperature limit
\cite{Vak85-90}.\,\,In the high-temperature limit, the expression
for the two-, \mbox{three-,} and four-particle structure factors are
reduced to the corresponding structure factors of the ideal Bose
gas.\,\,The two-particle structure factor obtained in the
approximation \textquotedblleft one sum over the wave
vector\textquotedblright\ also opens a way to finding the
temperature dependence of the first-sound velocity in liquid
$^{4}$He and to comparing it with experimental data.

\section{\boldmath$n$-Particle Structure\\ Factors for a Many-Boson System}

According to the definition, the $n$-particle structure factor equals
\[
S^{(n)}(\rho_{{\mathbf{q}}_{1}},...,\rho_{{\mathbf{q}}_{n}})=N^{n/2-1}%
\langle\rho_{{\mathbf{q}}_{1}}...\rho_{{\mathbf{q}}_{n}}\rangle,
\]
where $N$ is the number of particles, \mbox{$\rho_{q}=$}\linebreak
$=\frac{1}{\sqrt{N}}\sum _{j=1}^{N}e^{-i\mathbf{{qr_{j}}}}$ are
collective variables, and the notation $\langle...\rangle$ means the
averaging with the density matrix of interacting Bose
particles.\,\,In the calculations to follow, the density matrix
in the approximation \textquotedblleft two sums over the wave vector\textquotedblright%
\ with the factorized density matrix of the ideal Bose gas will be
used.\,\,This approximation involves the direct three- and
four-particle correlations, and the matrix itself looks like
\[
R(\rho|\rho^{\prime})=R_{N}^{0}(r|r^{\prime})P_{pr}(\rho|\rho^{\prime}%
)P(\rho|\rho^{\prime}),
\]
where $R_{N}^{0}(r|r^{\prime})$ is the density matrix for noninteracting Bose
particles, $P_{pr}(\rho|\rho^{\prime})$ a factor taking into account pair correlations, and $P(\rho|\rho^{\prime})$ a factor taking the direct
three- and four-particle correlations into account.\,\,In particular,%
 \[R_N^0(r'|r)=\frac{1}{N!}\left(\!\frac{m}{2\pi\beta\hbar^2}\!\right)^{\!\!3N/2}\times\]\vspace*{-7mm}
\[\times\sum\limits_Q
 \exp\left[-\frac{m}{2\beta\hbar^2}\sum\limits_{j=1}^N(r_j'-r_{Q_j})^2\right]\!\!,\]
where the summation over $Q$ means the summation over all
permutations of particle coordinates.\,\,The factor which makes
allowance for pair correlations looks like
\cite{Vak2004}\vspace*{-3mm}
\[
P_{pr}(\rho|\rho')=\exp\biggl[b_0+\sum_{\mathbf{q}\neq0}b_1(q){\rho_
{\mathbf{q}}}'\rho_{-\mathbf{q}}\,-
\]\vspace*{-7mm}
\[
-\,\frac{1}{2}\sum_{\mathbf{q}\neq0}b_2(q)\rho_
{\mathbf{q}}\rho_{-\mathbf{q}}\biggr]\!,
\]
where\vspace*{-3mm}
 \[
 b_0=-\beta E_0+\frac{1}{2}\sum_{{\mathbf
q}\neq0}\ln\left[\frac{\alpha_{q}\tanh\left(\!\frac{\beta
E_{q}}{2}\!\right)}{\tanh\left(\!\frac{\beta
\varepsilon_{q}}{2}\!\right)}\right]+
\]\vspace*{-6mm}
\[+\sum_{{\mathbf
q}\neq0}\ln\left(\!\frac{1-e^{-\beta\varepsilon_{q}}}{1-e^{-\beta
E_{q}}}\!\right)\!\!,\]\vspace*{-6mm}
\[
 b_1(q)=\frac{1}{2}\left(\!\frac{\alpha_q}{\sinh(\beta E_q)}-\frac{1}{\sinh(\beta\varepsilon_q)}\!\right)\!\!,\]\vspace*{-6mm}
\[
 b_2(q)=\frac{1}{2}\left(\alpha_q\coth(\beta E_q)-\coth(\beta\varepsilon_q)\right)\!,
\]\vspace*{-6mm}
\[
\alpha_q=\sqrt{1+{\frac{2N}{V}\nu_q}\left\slash\frac{\hbar^2q^2}{2m}\right.},
\]\vspace*{-6mm}
\[
E_q=\varepsilon_q\alpha_q=\frac{\hbar^2q^2}{2m}\alpha_q,
\]\vspace*{-6mm}
\[
\nu_{q}=\int e^{-i\mathbf{qr}}\Phi(r)d\mathbf{r}%
\]
is the Fourier coefficient of the pair interaction energy between
particles, and $\beta=1/T$ is the inverse temperature.\,\,An
expression for $P(\rho|\rho^{\prime})$ was presented in work
\cite{VakHryh1}.\,\,Its simplified version can be found in
Appendix~1.\,\,As a result, we obtain
%1
\[
\langle\rho_{{\mathbf q}_1}...\rho_{{\mathbf
q}_n}\rangle=\frac{1}{Z}\int d{\mathbf r}_1...\int d{\mathbf
r}_N\,\times
\]\vspace*{-7mm}
\[
\times\, R_N^0(r|r)P_{pr}(\rho|\rho)P(\rho|\rho) \rho_{{\mathbf
q}_1}...\rho_{{\mathbf q}_n}=
\]\vspace*{-5mm}
\[
=\frac{1}{Z}\int d{\mathbf r}_1...\int d{\mathbf
r}_NR_N^0(r|r)\rho_{{\mathbf q}_1}...\rho_{{\mathbf q}_n}\,\times
\]\vspace*{-5mm}
\[
\times\,\exp\Bigg[b_0-\frac{1}{2}\sum_{\mathbf{q}\neq0}\lambda_q\rho_
{\mathbf{q}}\rho_{-\mathbf{q}}+C_0+2\sum_{\mathbf{q}\neq0}C_2({\mathbf{q}})
\rho_{\mathbf{q}}\rho_{-{\mathbf q}}\,+
\]\vspace*{-5mm}
\[
+\frac{2}{\sqrt{N}}\mathop{\sum_{\mathbf{q}_1\neq0}\sum_{\mathbf{q}_2\neq0}\sum_{\mathbf{q}_3\neq0}}
\limits_{{\mathbf q}_1+{\mathbf q}_2+{\mathbf q}_3=0} C_3({\mathbf
q}_1,{\mathbf q}_2, {\mathbf q}_3)\rho_
{\mathbf{q}_1}\rho_{\mathbf{q}_2}\rho_{\mathbf{q}_3}\,+
\]%\vspace*{-5mm}
\begin{equation} \label{n_str_fact}
+\,\frac{2}{N}\sum\limits_{\mathbf{q}_1\neq0}\sum\limits_{\mathbf{q}_2\neq0}
C_4(\mathbf{q}_1,\mathbf{q}_2)
\rho_{\mathbf{q}_1}\rho_{-\mathbf{q}_1}
\rho_{\mathbf{q}_2}\rho_{-\mathbf{q}_2}\Bigg]\!,
\end{equation}
where%
 \[
 \lambda_q=2b_2(q)-b_1(q)=
 \]\vspace*{-7mm}
 \[
 =\alpha_{q}\tanh\left(\!\frac{\beta}{2}
E_{q}\!\right)-\tanh\left(\!\frac{\beta}{2}\varepsilon_{q}\!\right)\!\!.
\]

Explicit expressions for the quantities $C_{0}$,
$C_{2}({\mathbf{q}}_{1})$,
$C_{3}({\mathbf{q}}_{1},{\mathbf{q}}_{2},{\mathbf{q}}_{3})$, and
$C_{4}({\mathbf{q}}_{1},{\mathbf{q}}_{2})$ are given in
Appendix~2.\,\,They can be obtained using the data quoted in
Appendix~1.

\section{Pair Structure Factor}

In the case of pair structure factor ($n=2$), expression (\ref{n_str_fact})
can be rewritten in the form of a derivative with respect to the parameter
$\lambda_{q}$:
\[
\langle\rho_{{\mathbf q}}\rho_{-{\mathbf
q}}\rangle=\frac{d}{d\lambda_q}\ln\Biggl\{\!\int d{\mathbf
r}_1...\int d{\mathbf r}_N R_N^0(r|r)\,\times
\]\vspace*{-7mm}
\[\times\,\exp\Bigg[b_0-\frac{1}{2}\sum_{\mathbf{q}\neq0}\lambda_q\rho_
{\mathbf{q}}\rho_{-\mathbf{q}}\Bigg] P(\rho|\rho)\!\Biggr\}\!.
\]
In the adopted approximation \textquotedblleft two sums over the wave
vector\textquotedblright\, this expression can be
written as
follows:%
\[
\langle\rho_{{\mathbf q}}\rho_{-{\mathbf q}}\rangle
=\frac{d}{d\lambda_q}\ln\Biggl\{\!\int d{\mathbf r}_1...\int
d{\mathbf r}_NR_N^0(r|r)\,\times
\]\vspace*{-7mm}
\[\times\,\exp\Bigg[b_0-\frac{1}{2}\sum_{\mathbf{q}\neq0}\lambda_q\rho_
{\mathbf{q}}\rho_{-\mathbf{q}}\Bigg]\!\Biggr\}+\frac{d}{d\lambda_q}\ln\left\{\!\left\langle
P(\rho|\rho)\right\rangle\!\right\}\!.
\]
The expression for the first term was given in works
\cite{VP2008-09}.\,\,The
average $\left\langle P(\rho|\rho)\right\rangle $ looks like%
\[
\left\langle P(\rho|\rho)\right\rangle=
\]\vspace*{-7mm}
\[
=\!\frac{\int\!\! d{\bf r}_1...\int\!\! d{\bf r}_N R_N^0({\bf
r}_1,...,{\bf r}_N|{\bf r}_1,...,{\bf
r}_N)P_{pr}(\rho|\rho)P(\rho|\rho)}{\int\! d{\bf r}_1...\int d{\bf
r}_N R_N^0({\bf r}_1,...,{\bf r}_N|{\bf r}_1,...,{\bf
r}_N)P_{pr}(\rho|\rho)}.
\]
It can be obtained on the basis of work \cite{VakHryh2} as follows:%
\[
\label{P_ser} \left\langle
P(\rho|\rho)\right\rangle=\exp\biggl\{\!C_0+2\sum_{\mathbf{q}_1\neq0}C_2({\mathbf
q}_1)\frac{S_0( q_1)}{1+\lambda_{q_1}S_0(q_1)}\,+
\]\vspace*{-7mm}
\[
+\,\frac{2}{N}\!\!\sum_{\mathbf{q}_1\neq0}\sum_{\mathbf{q}_2\neq0}\!\!C_4({\mathbf
q}_1,{\mathbf q}_2)\frac{S_0(
q_1)}{1\!+\!\lambda_{q_1}S_0(q_1)}\frac{S_0(
q_2)}{1\!+\!\lambda_{q_2}S_0(q_2)}\,+
\]%\vspace*{-7mm}
\[
+\,\frac{2}{N}\mathop{\sum_{\mathbf{q}_1
\neq0}\sum_{\mathbf{q}_2\neq0}\sum_{\mathbf{q}_3\neq0}}\limits_{\mathbf{q}_1+\mathbf{q}_2+\mathbf{q}_3=0}
\frac{C_3({\mathbf q}_1,{\mathbf q}_2,{\mathbf q}_3)S_0^{(3)}(
{\mathbf q}_1,{\mathbf q}_2,{\mathbf
q}_3)}{\prod\limits_{i=1}^3[1+\lambda_{q_i}S_0(q_i)]}\,+
\]\vspace*{-7mm}
\[
+\,\frac{12}{N}\!\!\mathop{\sum_{\mathbf{q}_1\neq0}\sum_{\mathbf{q}_2\neq0}
\sum_{\mathbf{q}_3\neq0}}\limits_{\mathbf{q}_1+\mathbf{q}_2+\mathbf{q}_3=0}
\!\!C_3^2({\mathbf q}_1,{\mathbf q}_2,{\mathbf
q}_3)\!\prod\limits_{i=1}^3\!\frac{S_0(q_i)}{[1\!+\!\lambda_{q_i}S_0(q_i)]}\!\biggr\}\!.
\]

Therefore, using the explicit form for $\left\langle P(\rho|\rho)\right\rangle
$ and the results of works \cite{VP2008-09}, we obtain
%2
\[
S(q_1)=\left\langle\rho_
{\mathbf{q}_1}\rho_{-\mathbf{q}_1}\right\rangle=
\]\vspace*{-9mm}
\[=\frac{S_0(q_1)}{1+\lambda_{q_1}S_0(q_1)}-
\frac{1}{[1+\lambda_{q_1}S_0(q_1)]^2}\,\times
\]\vspace*{-7mm}
\[
\times\,\biggl(\!
\frac{1}{2N}\sum_{\mathbf{k}_2\neq0}\frac{\lambda_{k_2}S_0^{(4)}({\bf
q}_1,-{\bf q}_1, {\bf k}_2,-{\bf k}_2)}{1+\lambda_{k_2}S_0(k_2)}\,+
\]\vspace*{-7mm}
\[
+\,\frac{1}{2N}
\mathop{\sum_{\mathbf{k}_2\neq0}\!\sum_{\mathbf{k}_3\neq0}}
\limits_{{\mathbf q}_1+{\mathbf k}_2+{\mathbf
k}_3=0}\frac{\lambda_{k_2}}{1+\lambda_{k_2}S_0(k_2)}\frac{\lambda_{k_3}}{1+\lambda_{k_3}S_0(q_3)}\,\times
\]\vspace*{-7mm}
\[\times\left[S_0^{(3)}({\bf q}_1,{\bf k}_2, {\bf
k}_3)\right]^2 +4C_2({\mathbf{q}_1})S_0^2(q_1)\,+
\]\vspace*{-7mm}
\[
+\,\mathop{\sum_{\mathbf{k}_2\neq0}\sum_{\mathbf{k}_3\neq0}}
\limits_{{\mathbf q}_1+{\mathbf k}_2+{\mathbf
k}_3=0}\frac{C_3({\mathbf q}_1,{\mathbf k}_2, {\mathbf
k}_3)S_0^{(3)}({\bf q}_1,{\bf k}_2, {\bf
k}_3)}{[1+\lambda_{k_2}S_0(k_2)][1+\lambda_{k_3}S_0(k_3)]}\,+
\]\vspace*{-7mm}
\[
+\,\frac{8}{N} S_0^2(q_1)\! \sum_{\mathbf{k}_2\neq0}C_4({\mathbf
q}_1,{\mathbf k}_2)\frac{S_0(k_2)}{1+\lambda_{k_2}S_0(k_2)}\,+
\]\vspace*{-7mm}
\begin{equation}\label{S2}
+\,\frac{72}{N} S_0^2(q_1)\!\!\!
\mathop{\sum_{\mathbf{k}_2\neq0}\sum_{\mathbf{k}_3\neq0}}
\limits_{{\mathbf q}_1+{\mathbf k}_2+{\mathbf k}_3=0}
\frac{C_3^2({\mathbf q}_1,{\mathbf k}_2, {\mathbf
k}_3)S_0(k_2)S_0(k_3)}{[1\!+\!\lambda_{k_2}S_0(k_2)][1\!+\!\lambda_{k_3}S_0(k_3)]}\!\biggr)\!.
\end{equation}
Supposing the terms with a single sum to be small in comparison with the
quantity corresponding to the pair correlation approximation, the two-particle
structure factor can be written in the form%
\[
S(q_1)=\frac{S_0(q_1)}{1+(\lambda_{q_1}+\Pi_{q_1})S_0(q_1)},
\quad\Pi_{q_1}=\Pi_{q_1}^{np}+\Pi_{q_1}^{p}\!,
\]
where\vspace*{-3mm}
\[
\Pi_{q_1}^{np}=
\frac{1}{2NS_0^2(q_1)}\sum_{\mathbf{k}_2\neq0}\frac{\lambda_{k_2}S_0^{(4)}({\bf
q}_1,-{\bf q}_1, {\bf k}_2,-{\bf k}_2)}{1+\lambda_{k_2}S_0(k_2)}\,-
\]\vspace*{-7mm}
\[
-\,\frac{1}{2NS_0^2(q_1)}\!\!\!\mathop{\sum_{\mathbf{k}_2\neq0}\sum_{\mathbf{k}_3\neq0}}
\limits_{{\mathbf q}_1+{\mathbf k}_2+{\mathbf
k}_3=0}\!\!\!\frac{\lambda_{k_2}\lambda_{k_3}\left[S_0^{(3)}({\bf
q}_1,{\bf k}_2, {\bf
k}_3)\right]^2}{[1\!+\!\lambda_{k_2}S_0(k_2)][1\!+\!\lambda_{k_3}S_0(q_3)]}
\]
is the contribution of indirect three- and four-particle correlations, and
\[
\Pi_{q_1}^{p}=-4C_2({\mathbf{q}_1})-\frac{8}{N}\sum_{\mathbf{k}_2\neq0}
\frac{C_4({\mathbf q}_1,{\mathbf
k}_2)S_0(k_2)}{1+\lambda_{k_2}S_0(k_2)}\,-
\]\vspace*{-7mm}
\[
-\,\frac{12}{NS_0(q_1)}\!\!\!
\mathop{\sum_{\mathbf{k}_2\neq0}\sum_{\mathbf{k}_3\neq0}}
\limits_{{\mathbf q}_1+{\mathbf k}_2+{\mathbf k}_3=0}\!\!\!
\frac{C_3({\mathbf q}_1,{\mathbf k}_2, {\mathbf k}_3)S_0^{(3)}({\bf
q}_1,{\bf k}_2, {\bf
k}_3)}{[1\!+\!\lambda_{k_2}S_0(k_2)][1\!+\!\lambda_{k_3}S_0(k_3)]}\,-
\]\vspace*{-7mm}
\[
-\,\frac{72}{N}
\mathop{\sum_{\mathbf{k}_2\neq0}\sum_{\mathbf{k}_3\neq0}}
\limits_{{\mathbf q}_1+{\mathbf k}_2+{\mathbf k}_3=0}
\frac{C_3^2({\mathbf q}_1,{\mathbf k}_2, {\mathbf
k}_3)S_0(k_2)S_0(k_3)}{[1+\lambda_{k_2}S_0(k_2)][1+\lambda_{k_3}S_0(k_3)]}
\]
is the contribution of direct three- and four-particle correlations.

Expression (\ref{n_str_fact}) for the three-particle structure factor can be
presented in the form
\[
\left\langle\rho_
{\mathbf{q}_1}\rho_{\mathbf{q}_2}\rho_{\mathbf{q}_3}\right\rangle=-\frac{\sqrt{N}}{2}\frac{\delta
\ln\left\langle P(\rho|\rho)\right\rangle}{\delta C_3({\mathbf q}_1,
{\mathbf q}_2, {\mathbf q}_3)}.
\]
A direct calculation on the basis of the previous formula gives the
following result:
\[
\label{S3} S^{(3)}({\bf q}_1,{\bf q}_2, {\bf
q}_3)=\sqrt{N}\left\langle\rho_
{\mathbf{q}_1}\rho_{\mathbf{q}_2}\rho_{\mathbf{q}_3}\right\rangle=
\]\vspace*{-7mm}
\[
=\biggl\{\!\frac{S_0^{(3)}({\bf q}_1,{\bf q}_2,{\bf
q}_3)}{[1+\lambda_{q_1}S_0(q_1)][1+\lambda_{q_2}S_0(q_2)][1+\lambda_{q_3}S_0(q_3)]}\,+
\]\vspace*{-7mm}
\[
+\,\frac{12C_3({\mathbf q}_1,{\mathbf q}_2, {\mathbf
q}_3)S_0(q_1)S_0(q_2)S_0(q_3)}{[1+\lambda_{q_1}S_0(q_1)][1+
\lambda_{q_2}S_0(q_2)][1+\lambda_{q_3}S_0(q_3)]}\!\biggr\}\!.
\]

The irreducible four-particle structure factor takes the form
\[
S^{(4)}({\bf q}_1,-{\bf q}_1, {\bf q}_2,-{\bf q}_2)=
\]\vspace*{-9mm}
\[
=N\left[\left\langle\rho_{\mathbf{q}_1}\rho_{-\mathbf{q}_1}
\rho_{\mathbf{q}_2}\rho_{-\mathbf{q}_2}\right\rangle-\left\langle\rho_
{\mathbf{q}_1}\rho_{-\mathbf{q}_1}\right\rangle\left\langle\rho_
{\mathbf{q}_2}\rho_{-\mathbf{q}_2}\right\rangle\right]\!.
\]
The average $\left\langle \rho_{\mathbf{q}_{1}}\rho_{-\mathbf{q}_{1}%
}\right\rangle $ was found earlier, and we have to calculate $\left\langle
\rho_{\mathbf{q}_{1}}\rho_{-\mathbf{q}_{1}}\rho_{\mathbf{q}_{2}}%
\rho_{-\mathbf{q}_{2}}\right\rangle $.\,\,Again, on the basis of
formula
(\ref{n_str_fact}), it can be shown that%
\[
\left\langle\rho_{\mathbf{q}_1}\rho_{-\mathbf{q}_1}
\rho_{\mathbf{q}_2}\rho_{-\mathbf{q}_2}\right\rangle=\frac{1}{I_{\lambda}}
\frac{d^2 I_{\lambda}}{d\lambda_{q_1}d\lambda_{q_2}},
\]
where\vspace*{-3mm}
\[
I_{\lambda}=\int d{\mathbf r}_1...\int d{\mathbf r}_N
R_N^0(r|r)\,\times
\]\vspace*{-7mm}
\[
\times\,\exp\left[b_0-\frac{1}{2}\sum\limits_{\mathbf{q}\neq0}\lambda_q\rho_
{\mathbf{q}}\rho_{-\mathbf{q}}\right]P(\rho|\rho).
\]
Then,%
\[
\left\langle\rho_{\mathbf{q}_1}\rho_{-\mathbf{q}_1}
\rho_{\mathbf{q}_2}\rho_{-\mathbf{q}_2}\right\rangle-\left\langle\rho_
{\mathbf{q}_1}\rho_{-\mathbf{q}_1}\right\rangle\left\langle\rho_
{\mathbf{q}_2}\rho_{-\mathbf{q}_2}\right\rangle=
\]\vspace*{-7mm}
\[
=\frac{d^2}{d\lambda_{q_1}d\lambda_{q_2}}\ln\Biggl\{\!\int d{\mathbf
r}_1...\int d{\mathbf r}_N R_N^0(r|r)\,\times
\]\vspace*{-7mm}
\[\times\,\exp\Bigg[b_0-\frac{1}{2}\sum_{\mathbf{q}\neq0}\lambda_q\rho_
{\mathbf{q}}\rho_{-\mathbf{q}}\Bigg]\!\Biggr\}
+\frac{d^2\ln\left\langle
P(\rho|\rho)\right\rangle}{d\lambda_{q_1}d\lambda_{q_2}}.
\]
The first term in the expression above was also found earlier
\cite{VP2008-09}.\,\,The second one is easy to calculate taking  the
explicit expression for $\left\langle P(\rho|\rho)\right\rangle$
into account.\,\,As a result, we obtain
\[
\label{S4} S^{(4)}({\bf q}_1,-{\bf q}_1, {\bf q}_2,-{\bf
q}_2)=\frac{1}{[1+\lambda_{q_1}S_0(q_1)]^2}\,\times\]\vspace*{-7mm}
\[
\times\,\frac{1}{[1+\lambda_{q_2}S_0(q_2)]^2}
\Bigg\{\!S_0^{(4)}({\bf q}_1,-{\bf q}_1, {\bf q}_2,-{\bf q}_2)\,-
\]\vspace*{-7mm}
\[
-\,\frac{2\lambda_{|{\mathbf q}_1+{\mathbf
q}_2|}\left[S_0^{(3)}({\bf q}_1,{\bf q}_2, -{\bf q}_1-{\bf
q}_2)\right]^2}{1+\lambda_{|{\mathbf q}_1+{\mathbf
q}_2|}S_0(|{\mathbf q}_1+{\mathbf q}_2|)}\,+\]\vspace*{-7mm}
\[
+\,48 S_0(q_1) S_0(q_2)S_0^{(3)}({\bf q}_1,{\bf q}_2, -{\bf
q}_1-{\bf q}_2)\,\times\]\vspace*{-7mm}
\[\times\,
\frac{C_3({\mathbf q}_1,{\mathbf q}_2, -{\mathbf q}_1-{\mathbf
q}_2)}{1+\lambda_{{|{\mathbf q}_1+{\mathbf q}_2|}}S_0(|{\mathbf
q}_1+{\mathbf q}_2|)}\,+\]\vspace*{-7mm}
\[+\,
16 S_0^2(q_1) S_0^2(q_2)\Bigg[C_4({\mathbf q}_1,{\mathbf q}_2)\,+
\]\vspace*{-7mm}
\[+\,
18\frac{C_3^2({\mathbf q}_1,{\mathbf q}_2, -{\mathbf q}_1-{\mathbf
q}_2)S_0( |{\bf q}_1+{\bf q}_2|)}{1+\lambda_{|{\mathbf q}_1+{\mathbf
q}_2|}S_0(|{\mathbf q}_1+{\mathbf q}_2|)}\Bigg]\!\Bigg\}\!.
\]

\section{Two-, Three-, and Four-Particle Structure Factors in the
Low-Temperature Limit}

In the low-temperature limit, the pair and three-particle structure
factors equal unity, and the irreducible four-particle one equals
zero.\,\,Their derivatives with respect to the inverse temperature
vanish in this limit.\,\,One
may verify it directly by analyzing the corresponding expressions.%
\[
S_0(q)=1,\quad\frac{\partial S_0(q)}{\partial\beta}=0,
\]\vspace*{-7mm}
\[
S_0^{(3)}({\bf q}_1,{\bf q}_2, {\bf q}_3)=1,\quad\frac{\partial
S_0^{(3)}({\bf q}_1,{\bf q}_2, {\bf q}_3)}{\partial\beta}=0,
\]\vspace*{-7mm}
\[
S_0^{(4)}\!({\bf q}_1,\!-{\bf q}_1, {\bf q}_2,\!-{\bf
q}_2)=0,\frac{\partial S_0^{(4)}\!({\bf q}_1,\!-{\bf q}_1, {\bf
q}_2,\!-{\bf q}_2)}{\partial\beta}=0.
\]
A straightforward verification also demonstrates that
\[
\lim_{\beta\rightarrow\infty}C_2({\mathbf
q}_1)=\frac{1}{2}a_2({\mathbf q}_1),
\]\vspace*{-7mm}
\[
\lim_{\beta\rightarrow\infty}C_3({\mathbf q}_1,{\mathbf q}_2,
{\mathbf q}_3)=\frac{1}{6}a_3({\mathbf q}_1,{\mathbf q}_2, {\mathbf
q}_3),
\]\vspace*{-7mm}
\[
\lim_{\beta\rightarrow\infty}C_4({\mathbf q}_1,{\mathbf q}_2)
=\frac{1}{8}a_4({\mathbf q}_1,-{\mathbf q}_1,{\mathbf q}_2,-{\mathbf
q}_2),
\]
where the quantities $a_{2}({\mathbf{q}}_{1})$, $a_{3}{({\mathbf{q}}%
_{1},{\mathbf{q}}_{2},{\mathbf{q}}_{3})}$, and $a_{4}{({\mathbf{q}}%
_{1},-{\mathbf{q}}_{1},{\mathbf{q}}_{2},-{\mathbf{q}}_{2})}$ are the known
expressions \cite{VakUhn79-80} and look like%
\[
a_2({\mathbf q}_1)=\frac{1}{N}\sum_{{\mathbf q}_2\neq0}
\biggl[\frac{q_2^2}{2q_1^2\alpha_{q_1}} a_4{({\mathbf q}_1,-{\mathbf
q}_1,{\mathbf q}_2,-{\mathbf q}_2)}\,+
\]\vspace*{-7mm}
\[+\,
\frac{({\mathbf q}_2,{\mathbf q}_1+{\mathbf q}_2)}
{q_1^2\alpha_{q_1}}a_3{({\mathbf q}_1,{\mathbf q}_2,-{\mathbf
q}_1-{\mathbf q}_2)} \!\biggr]\!,
\]\vspace*{-7mm}
\[
a_3{({\mathbf q}_1,{\mathbf q}_2,{\mathbf
q}_3)}=-\frac{\sum\limits_{1\leq i<j\leq3} {({\mathbf q}_i{\mathbf
q}_j)}(\alpha_{q_i}-1)(\alpha_{q_j}-1)}{2\sum\limits_{j=1}^3
{\mathbf q}_j^2\alpha_{q_j}},
\]\vspace*{-7mm}
\[
a_4{({\mathbf q}_1,-{\mathbf q}_1,{\mathbf q}_2,-{\mathbf
q}_2)}=\frac{1}{q_1^2\alpha_{q_1}+q_2^2\alpha_{q_2}}\,\times
\]\vspace*{-7mm}
\[
\times\,\bigl\{\!({\mathbf q}_1+{\mathbf q}_2)^2 a_3^2({\mathbf
q}_1+{\mathbf q}_2,-{\mathbf q}_1,-{\mathbf q}_2)\,+
\]\vspace*{-9mm}
\[+\,({\mathbf
q}_1-{\mathbf q}_2)^2 a_3^2({\mathbf q}_1-{\mathbf q}_2,-{\mathbf
q}_1,{\mathbf q}_2)\,-
\]\vspace*{-9mm}
\[-\,[{({\mathbf q}_1,{\mathbf q}_2+ {\mathbf
q}_1)}(\alpha_{q_1}-1)+{({\mathbf q}_2,{\mathbf q}_1+ {\mathbf
q}_2)}(\alpha_{q_2}-1)]\,\times\]\vspace*{-9mm}
\[
\times\, a_3{({\mathbf q}_1+{\mathbf q}_2,-{\mathbf q}_1,-{\mathbf
q}_2)}\,-\]\vspace*{-9mm}
\[-\,[{({\mathbf q}_1,{\mathbf q}_1- {\mathbf
q}_2)}(\alpha_{q_1}-1)+{({\mathbf q}_2,{\mathbf q}_2- {\mathbf
q}_1)}(\alpha_{q_2}-1)]\,\times
\]\vspace*{-9mm}
\[
\times\, a_3{({\mathbf q}_1-{\mathbf q}_2,-{\mathbf q}_1,{\mathbf
q}_2)}\!\bigr\}\!.
\]

Taking the aforesaid into account, we obtain the following expression for the
pair structure factor in the low-temperature limit:
%3
\[
S(q_1)=\frac{1}{\alpha^2_{q_1}}\Biggl[\alpha_{q_1}+\frac{1}{2N}\!\!\!
\mathop{\sum_{\mathbf{k}_2\neq0}\sum_{\mathbf{k}_3\neq0}}
\limits_{{\mathbf q}_1+{\mathbf k}_2+{\mathbf
k}_3=0}\frac{\alpha_{k_2}-1}{\alpha_{k_2}}\,\times
\]\vspace*{-7mm}
\[
\times\,\frac{\alpha_{k_3}-1}{\alpha_{k_3}}+2a_2({\mathbf{q}_1})+\frac{2}{N}
\mathop{\sum_{\mathbf{k}_2\neq0}\sum_{\mathbf{k}_3\neq0}}
\limits_{{\mathbf q}_1+{\mathbf k}_2+{\mathbf
k}_3=0}\frac{a_3({\mathbf q}_1,{\mathbf k}_2, {\mathbf
k}_3)}{\alpha_{k_2}\alpha_{k_3}}\,+
\]\vspace*{-7mm}
\begin{equation}\label{S2_T0}
+\,\frac{1}{N}\!\! \sum_{\mathbf{k}_2\neq0}\frac{a_4({\mathbf
q}_1,{\mathbf k}_2)}{\alpha_{k_2}}\!+\!\frac{2}{N}\!\!
\mathop{\sum_{\mathbf{k}_2\neq0}\sum_{\mathbf{k}_3\neq0}}
\limits_{{\mathbf q}_1+{\mathbf k}_2+{\mathbf k}_3=0}
\!\!\!\frac{a_3^2({\mathbf q}_1,{\mathbf k}_2, {\mathbf
k}_3)}{\alpha_{k_2}\alpha_{k_3}}\!\Biggr]\!,
\end{equation}
where $a_{4}({\mathbf{q}}_{1},{\mathbf{k}}_{2})$ is an abbreviated notation
for the quantity $a_{4}{({\mathbf{q}}_{1},-{\mathbf{q}}_{1},{\mathbf{k}}%
_{2},-{\mathbf{k}}_{2})}$.\,\,In the adopted approximation
\textquotedblleft two sums over the wave vector\textquotedblright\,
the structure factor can be written in the form similar to that in
work \cite{Vak85-90},
\[
S(q)=1/[1-2a_{2}(q)-\Sigma(q)],
\]
where\vspace*{-3mm}
\[
\label{S2_T0_1_sigma} \Sigma(q)=\frac{1}{N}
\sum_{\mathbf{k}\neq0}\frac{a_4({\mathbf k},-{\mathbf k},{\mathbf
q},-{\mathbf q})}{1-2\tilde{a}_2(k)}+\frac{2}{N} \times
\]\vspace*{-7mm}
\[
\times\!\!\!\!\mathop{\sum_{\mathbf{k}_1\neq0}\sum_{\mathbf{k}_2\neq0}}
\limits_{{\mathbf q}+{\mathbf k}_1+{\mathbf
k}_2=0}\!\!\!\frac{\left\{\!\tilde{a}_2(k_1\!)\tilde{a}_2(k_2\!)\!+\!a_3({\mathbf
q},{\mathbf k}_1\!, {\mathbf k}_2)[1\!+\!a_3({\mathbf q},{\mathbf
k}_1\!, {\mathbf
k}_2)]\!\right\}}{(1-2\tilde{a}_2(k_1))(1-2\tilde{a}_2(k_2))},
\]\vspace*{-7mm}
\[\label{a2t}
\tilde{a}_2({ q}_1)=-\frac{1}{2}(\alpha_{q_1}-1)\,+
\]\vspace*{-7mm}
\[
+\,\frac{1}{N}\sum_{{\mathbf q}_2\neq0}
\biggl[\frac{q_2^2}{2q_1^2\alpha_{q_1}} a_4{({\mathbf q}_1,-{\mathbf
q}_1,{\mathbf q}_2,-{\mathbf q}_2)}\,+
\]\vspace*{-7mm}
\[+\,\frac{({\mathbf
q}_2,{\mathbf q}_1+{\mathbf q}_2)} {q_1^2\alpha_{q_1}}a_3{({\mathbf
q}_1,{\mathbf q}_2,-{\mathbf q}_1-{\mathbf q}_2)} \!\biggr]\!.
\]

Analogously, we obtain the following expressions for the three- and
four-particle structure factors in the low-temperature limit:
\[
\label{S3_T0} S^{(3)}({\bf q}_1,{\bf q}_2, {\bf
q}_3)=\frac{1}{\alpha_{q_1}\alpha_{q_2}\alpha_{q_3}}\left\{\!
1+2a_3({\mathbf q}_1,{\mathbf q}_2, {\mathbf q}_3)\!\right\}\!,
\]\vspace*{-7mm}
\[
\label{S4_T0} S^{(4)}({\bf q}_1,-{\bf q}_1, {\bf q}_2,-{\bf q}_2)=
\frac{2}{\alpha^2_{q_1}\alpha^2_{q_2}}\biggl[-\frac{\alpha_{|{\mathbf
q}_1+{\mathbf q}_2|}-1}{\alpha_{|{\mathbf q}_1+{\mathbf q}_2|}}\,+
\]\vspace*{-7mm}
\[+\,\frac{4}{\alpha_{|{\mathbf q}_1+{\mathbf q}_2|}}a_3({\mathbf
q}_1,{\mathbf q}_2, -{\mathbf q}_1-{\mathbf q}_2)\,+
\]\vspace*{-7mm}
\[+\,a_4({\mathbf q}_1,{\mathbf q}_2)+
\frac{4}{\alpha_{|{\mathbf q}_1+{\mathbf q}_2|}}a_3^2({\mathbf
q}_1,{\mathbf q}_2, -{\mathbf q}_1-{\mathbf q}_2)\!\biggr]\!.
\]

\section{Two-, Three-, and Four-Particle Structure Factors in the
High-Temperature Limit}

Using the explicit expressions for the quantities $\overline{C}_{2}%
(\mathbf{q_{1}})$, $\overline{C}_{2}^{0}(\mathbf{q_{1}})$, $\overline{C}%
_{3}(\mathbf{q_{1}},\mathbf{q_{2}},\mathbf{q_{3}})$, $\overline{C}_{3}%
^{0}(\mathbf{q_{1}},\mathbf{q_{2}},\mathbf{q_{3}})$, $\overline{C}%
_{4}(\mathbf{q_{1}},\mathbf{q_{2}})$, and $\overline{C}_{4}^{0}(\mathbf{q_{1}%
},\mathbf{q_{2}})$ (see Appendix~2), we can easily obtain that, in
the high-temperature limit ($T\rightarrow\infty$ or
$\beta\rightarrow0$),
\[
\lim_{\beta\rightarrow0}\overline{C}_2({\bf
q_1})=\lim_{\beta\rightarrow0}\overline{C}_2^0({\bf
q_1})=\frac{1}{16N}\!\!\!\!\mathop{\sum_{\mathbf{q}_2\neq0}\sum_{\mathbf{q}_3\neq0}}
\limits_{{\mathbf q}_1+{\mathbf q}_2+{\mathbf q}_3=0}\!\!\!
\frac{q_1^2({\bf q}_2{\bf q}_3)}{q_2^2q_3^2},
\]\vspace*{-7mm}
\[
\lim_{\beta\rightarrow0}\overline{C}_3({\bf q_1},{\bf q_2},{\bf
q_3})=\lim_{\beta\rightarrow0}\overline{C}_3^0({\bf q_1},{\bf
q_2},{\bf q_3})=\frac{1}{12},
\]\vspace*{-7mm}
\[
\lim_{\beta\rightarrow0}\overline{C}_4({\bf q_1},{\bf
q_2})=\lim_{\beta\rightarrow0}\overline{C}_4^0({\bf q_1},{\bf
q_2})=\frac{1}{8},\]
so that
\[
\lim_{\beta\rightarrow0}C_2({\bf
q_1})\!=\!\lim_{\beta\rightarrow0}C_3({\bf q_1},{\bf q_2},{\bf
q_3})\!=\!\lim_{\beta\rightarrow0}C_4({\bf q_1},{\bf q_2})=0.
\]
Therefore, in the high-temperature limit, the two-, three-, and
four-particle structure factors for a many-boson system transform
into the corresponding expressions for the ideal Bose gas:
\[
\lim_{\beta\rightarrow0}S(q)=S_0(q),
\]\vspace*{-7mm}
\[
\lim_{\beta\rightarrow0}S^{(3)}({\bf q_1},{\bf q_2},{\bf
q_3})=S_0^{(3)}({\bf q_1},{\bf q_2},{\bf q_3}),
\]\vspace*{-7mm}
\[
\lim_{\beta\rightarrow0}S^{(4)}({\bf q_1},-{\bf q_1},{\bf q_2},-{\bf
q_2})=S_0^{(4)}({\bf q_1},-{\bf q_1},{\bf q_2},-{\bf q_2}).
\]

\section{Numerical Calculations}

The numerical calculation of the two-particle structure factor
(\ref{S2}) will be carried out taking the effective mass into
account \cite{HP2014}.\,\,In order to not exceed the calculation
accuracy, the effective mass will be used only in the terms that
reproduce the pair correlation approximation.\,\,At the same time,
the expressions containing the sum over the wave vector will contain
a \textquotedblleft bare\textquotedblright\ mass.\,\,Here, the
following remark is worth making: in the structure factors of the
ideal Bose gas which enter the expressions with a sum over the wave
vector, the effective mass is used only
to shift the critical point owing to  the activity renormalization$\ z_{0}%
=\exp[\beta\mu]$, where $\mu$ is the chemical potential.\,\,The
introduction of effective mass makes it possible to avoid infra-red
divergences in the non-renormalized four-particle structure factor
of the ideal Bose gas.

To calculate the quantities with a single sum over the wave vector, we should
change from summation to integration according to the well-known rule
\cite{Vstup}%
\[
\sum_{\mathbf{k}}=\frac{V}{(2\pi)^{3}}\int d\mathbf{k}.
\]
After the corresponding transformations and the required changes in variables, we
obtain the following rule for the change from summation to integration in our
case:%
\[
\frac{1}{N}%
\mathop{\sum_{\mathbf{k}_2\neq0}\sum_{\mathbf{k}_3\neq0}}\limits_{{\mathbf{q}%
}_{1}+{\mathbf{k}}_{2}+{\mathbf{k}}_{3}=0}=\frac{1}{4\pi^{2}\rho q_{1}}%
\int\limits_{0}^{\infty}k_{2}dk_{2}\int\limits_{|q_{1}-k_{2}|}^{|q_{1}+k_{2}|}%
k_{3}dk_{3},
\]
where $\rho$ is the equilibrium density in the Bose system.\,\,For
such quantum liquid as $^{4}$He, the latter parameter equals
$\rho=0.02185$~{\AA} $^{-3}$ \cite{DonBar}.\,\,The next step
consists in calculating the quantities $\alpha_{q}$ with the use of
a pair structure factor at the zero temperature extrapolated on the
basis of experimental data.\,\,The corresponding information is
taken from work \cite{VBR}.

%Fig1
\begin{figure}
\vskip1mm
\includegraphics[width=\column]{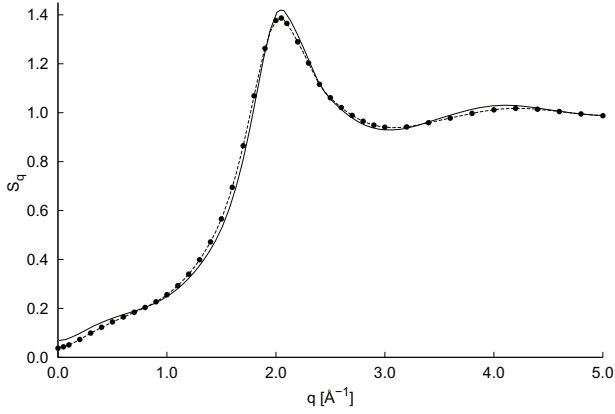}
\vskip-4mm\caption{Structure factor of liquid $^{4}$He at the
temperature $T=1.0$~K  }
\end{figure}

Now, let us rewrite Eq.~(\ref{S2_T0}) in the form
%4
\[
 \frac{1}{\alpha_{q_1}}=S^{\exp}(q_1)\,-
\]\vspace*{-7mm}
\[-\,\frac{1}{\alpha^2_{q_1}}\biggl[\frac{1}{2N}\mathop{\sum_{\mathbf{k}_2\neq0}\sum_{\mathbf{k}_3\neq0}}
\limits_{{\mathbf q}_1+{\mathbf k}_2+{\mathbf
k}_3=0}\frac{\alpha_{k_2}-1}{\alpha_{k_2}}\frac{\alpha_{k_3}-1}{\alpha_{k_3}}\,+
\]\vspace*{-7mm}
\[
+\,2a_2({\mathbf{q}_1})+\frac{2}{N}
\mathop{\sum_{\mathbf{k}_2\neq0}\sum_{\mathbf{k}_3\neq0}}
\limits_{{\mathbf q}_1+{\mathbf k}_2+{\mathbf
k}_3=0}\frac{a_3({\mathbf q}_1,{\mathbf k}_2, {\mathbf
k}_3)}{\alpha_{k_2}\alpha_{k_3}}\,+
\]\vspace*{-7mm}
\begin{equation}\label{renormaq}
+\, \frac{1}{N}\!\! \sum_{\mathbf{k}_2\neq0}\frac{a_4({\mathbf
q}_1,{\mathbf k}_2)}{\alpha_{k_2}}+\frac{2}{N}\!\!\!
\mathop{\sum_{\mathbf{k}_2\neq0}\sum_{\mathbf{k}_3\neq0}}
\limits_{{\mathbf q}_1+{\mathbf k}_2+{\mathbf k}_3=0}\!\!\!
\frac{a_3^2({\mathbf q}_1,{\mathbf k}_2, {\mathbf
k}_3)}{\alpha_{k_2}\alpha_{k_3}}\!\biggr]\!.
\end{equation}
This is an iterative equation for $\alpha_{q}$.\,\,In the zero-order
approximation, we have $\alpha_{q}=1/S^{\exp}(q)$.\,\,Substituting
this $\alpha_{q}$-value into the right-hand side of equality
(\ref{renormaq}), we obtain the $\alpha_{q}$-value in the first
approximation, and so forth.\,\,However, this iteration process does
not converge, which is most likely connected with an insufficient
number of terms in the series expansion for the structure factor
(\ref{S2_T0}).\,\,Therefore, the consideration will be confined only
to the zero-order approximation for~$\alpha_{q}$.

%Fig2
\begin{figure}
\vskip1mm
\includegraphics[width=\column]{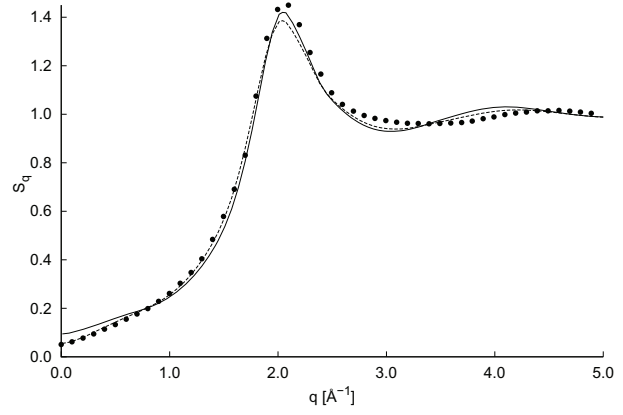}
\vskip-3mm\caption{The same as in Fig.~1, but at $T=1.38$~K  }
\end{figure}

%Fig3
\begin{figure}
\vskip7mm
\includegraphics[width=\column]{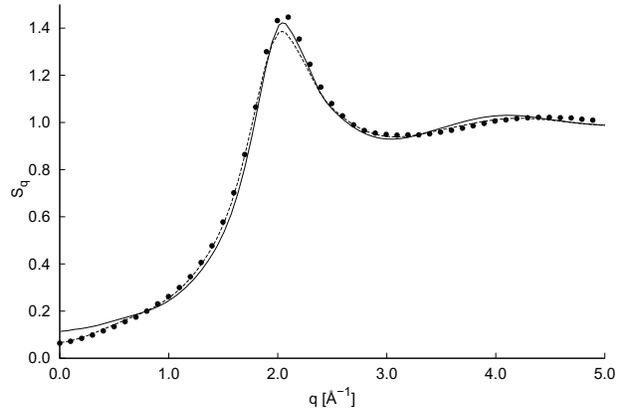}
\vskip-3mm\caption{The same as in Fig.~1, but at $T=1.67$~K  }
\end{figure}

%Fig4
\begin{figure}
\vskip7mm
\includegraphics[width=\column]{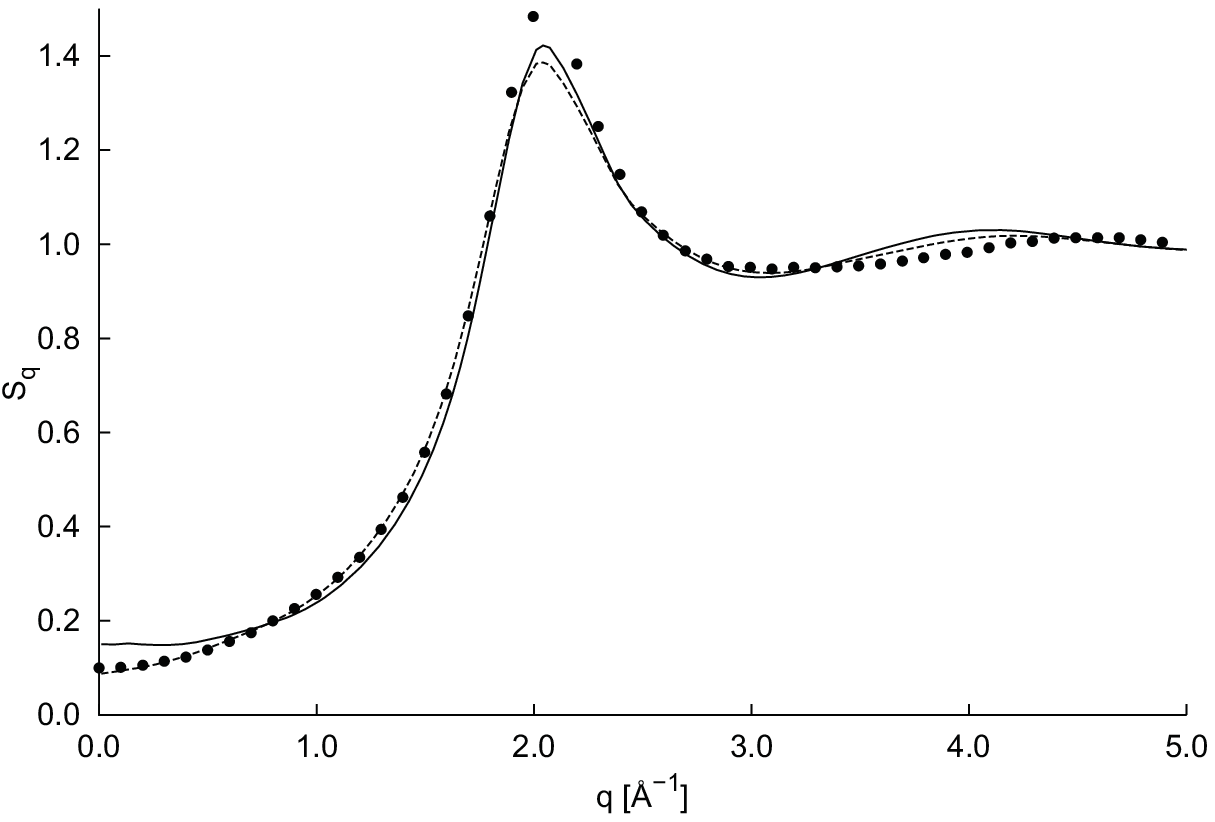}
\vskip-3mm\caption{The same as in Fig.~1, but at $T=2.2$~K  }
\end{figure}

%Fig5
\begin{figure}
\vskip1mm
\includegraphics[width=\column]{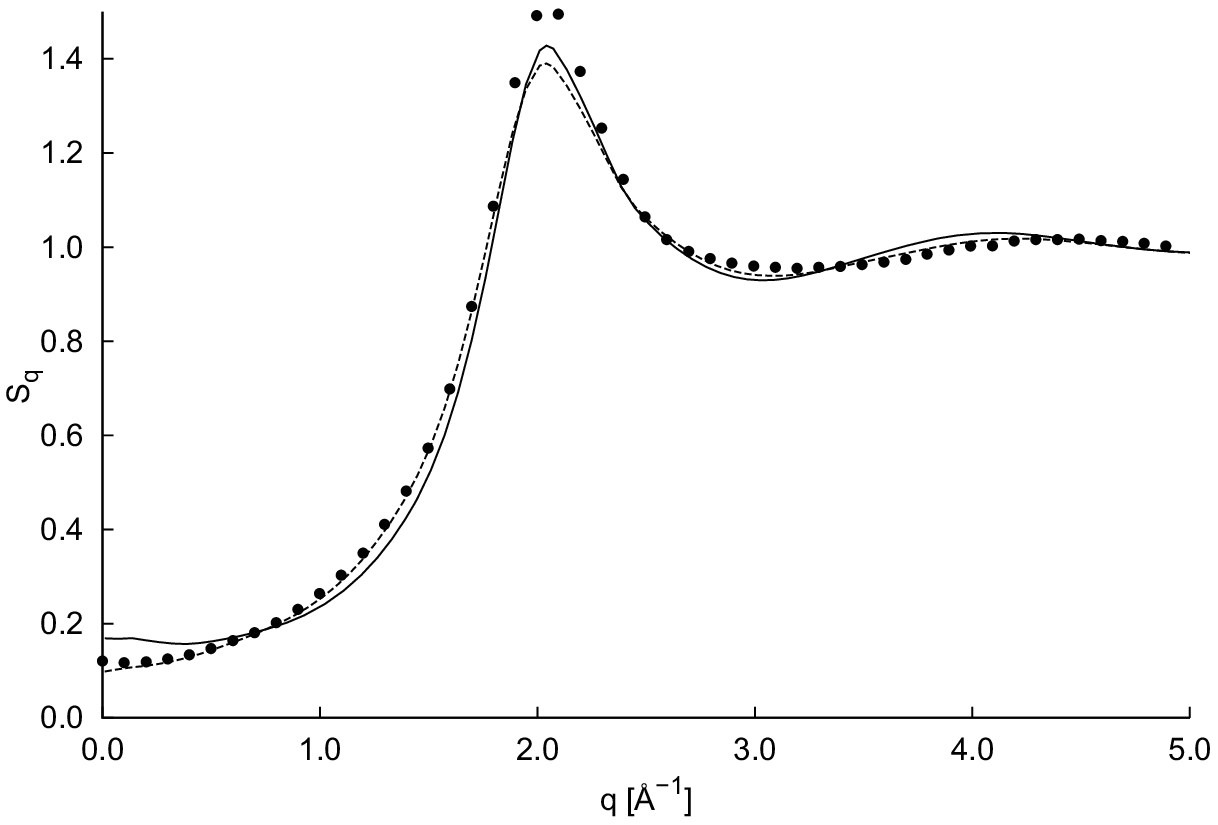}
\vskip-3mm\caption{The same as in Fig.~1, but at $T=2.5$~K  }
\end{figure}

%Fig6
\begin{figure}
\vskip7mm
\includegraphics[width=\column]{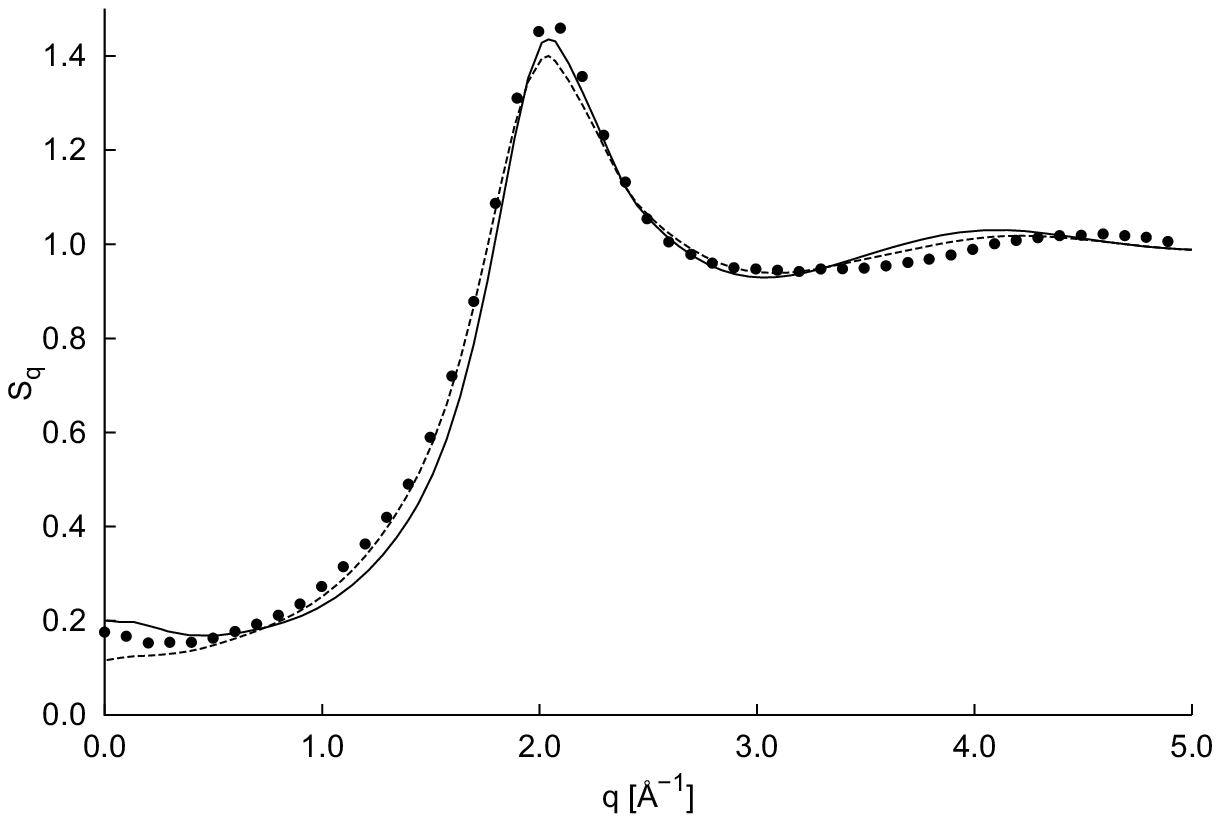}
\vskip-3mm\caption{The same as in Fig.~1, but at $T=3.0$~K  }
\end{figure}

%Fig7
\begin{figure}
\vskip7mm
\includegraphics[width=\column]{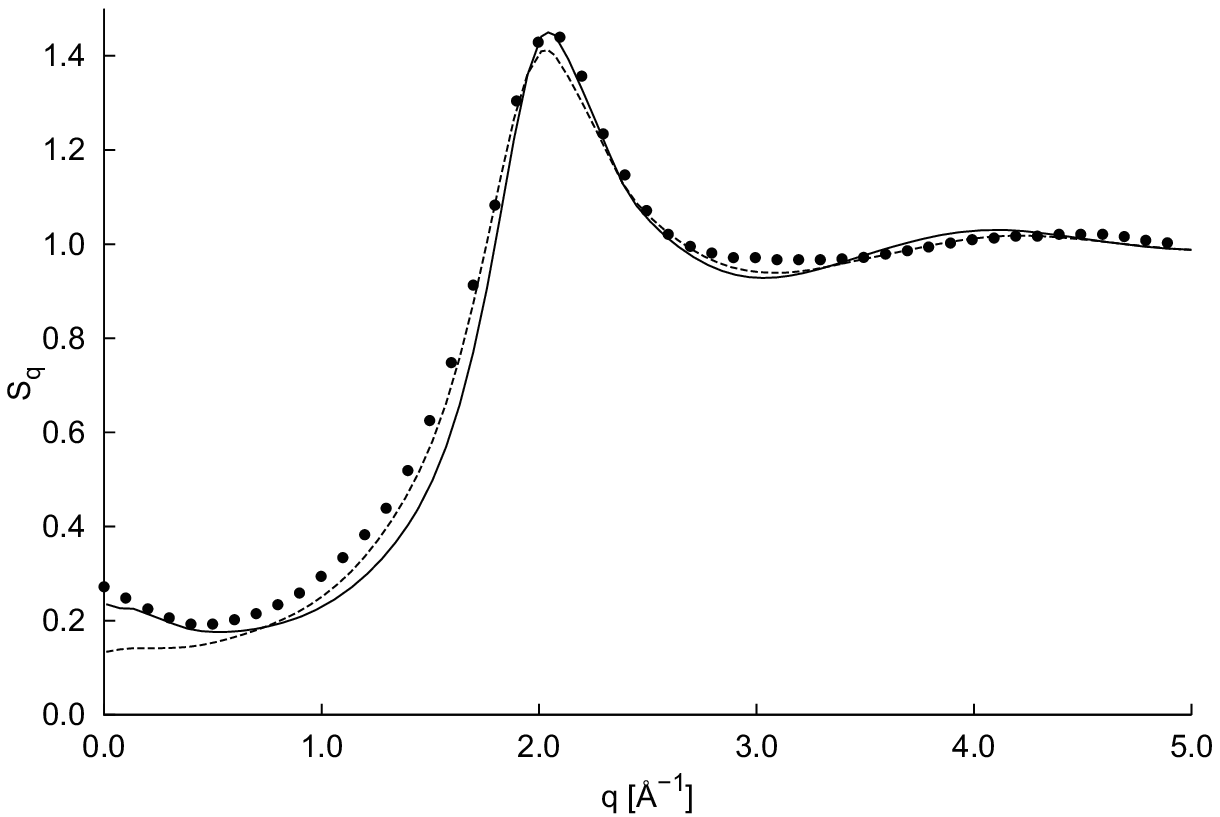}
\vskip-3mm\caption{The same as in Fig.~1, but at $T=3.5$~K  }
\end{figure}

%Fig8
\begin{figure}
\vskip1mm
\includegraphics[width=\column]{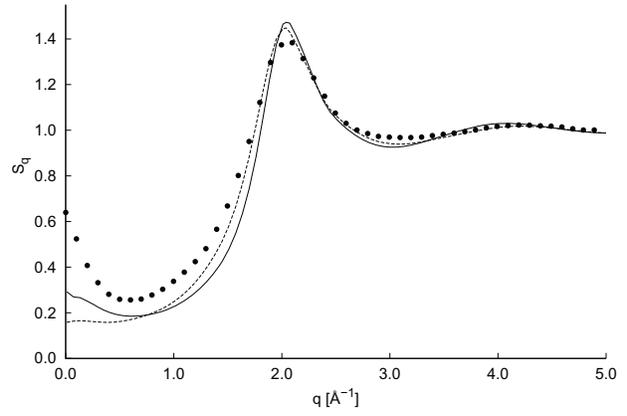}
\vskip-3mm\caption{ The same as in Fig.~1, but at $T=4.24$~K }
\end{figure}

The results of numerical calculations for temperatures of 1.0, 1.38,
1.67, 2.2, 2.5, 3.0, 3.5, and 4.24~K are exhibited in Figs.~1 to 8,
respectively. Experimental data for the structure factor at those
temperatures were taken from works \cite{Svensson, Robkoff}.\,\,In
the presented figures, the solid curves correspond to the structure
factor calculated with regard for the direct three- and
four-particle correlations, the dashed curves to the pair
correlation approximation, and the circles to the experimental
structure factor values.

\section{Conclusions}

In this work, expressions for the two-, three-, and four-particle
structure factors in a wide temperature interval were found in the
approximation \textquotedblleft one sum over the wave
vector\textquotedblright\ with regard for the direct three- and
four-particle correlations.\,\,In the low-temperature limit, the
expression obtained for the two-particle structure factor transforms
into the well-known one \cite{Vak85-90}.\,\,The same is valid for
the high-temperature limit.

The derived expressions are rather cumbersome. They were analyzed,
by using numerical methods, and graphic representations of the pair
structure factor at various temperatures of liquid $^{4}$He were
plotted.\,\,The calculation of the internal energy and the
determination of the temperature dependence of the first-sound
velocity in the many-boson system will be a subject of our next
papers.

\subsubsection*{APPENDIX 1}
{\footnotesize
\[
P(\rho|\rho')=\exp\biggl[\frac{c_0}{N}+\sum_{\mathbf{q}_1\neq0}
\sum_{i_1=0}^1 \sum_{j_1=0}^1 c_2\left(1^{j1},-1^{i_1}\right)\times
\]
\[
\times
\left(\rho_{\mathbf{q}_1}^{j_1}\rho_{-\mathbf{q}_1}^{i_1}+\rho_{\mathbf{q}_1}^{1-j_1}\rho_{-\mathbf{q}_1}^{1-i_1}\right)+
\]
\[
+\,\frac{1}{\sqrt{N}}\mathop{\sum_{\mathbf{q}_1\neq0}\sum_{\mathbf{q}_2\neq0}\sum_{\mathbf{q}_3\neq0}}
\limits_{\mathbf{q}_1+\mathbf{q}_2+\mathbf{q}_3\neq0}\sum_{i_1,i_2,i_3=0}^1
c_3(1^{i_1},2^{i_2},3^{i_3})\,\times
\]
\[
\times\left(\rho_
{\mathbf{q}_1}^{i_1}\rho_{\mathbf{q}_2}^{i_2}\rho_{\mathbf{q}_3}^{i_3}+\rho_
{\mathbf{q}_1}^{1-i_1}\rho_{\mathbf{q}_2}^{1-i_2}\rho_{\mathbf{q}_3}^{1-i_3}\!\right)+
\]
\[
+\,\frac{1}{N}\sum_{\mathbf{q}_1\neq0}\sum_{\mathbf{q}_2\neq0}
\sum_{i_1,i_2=0}^1 \sum_{j_1,j_2=0}^1
c_4(1^{j_1},-1^{i_1},2^{j_2},-2^{i_2})\,\times
\]
\[
\times\,\left(\rho_{\mathbf{q}_1}^{j_1}\rho_{-\mathbf{q}_1}^{i_1}
\rho_{\mathbf{q}_2}^{j_2}\rho_{-\mathbf{q}_2}^{i_2}+\rho_{\mathbf{q}_1}^{1-j_1}\rho_{-\mathbf{q}_1}^{1-i_1}
\rho_{\mathbf{q}_2}^{1-j_2}\rho_{-\mathbf{q}_2}^{1-i_2}\!\right)\!\biggr]\!.
\]

The superscripts $i_{1}$, $i_{2}$, $i_{3}$, $j_{1},$ and $j_{2}$ at
the quantities $\rho_{\mathbf{q}}$ can acquire values of 0 or 1,
namely, 0 means the absence of the prime, and 1 its presence.\,\,The
notation $c_{2}\left(
1^{j1},-1^{i_{1}}\right)  $ stands for $c_{2}\left(  \rho_{q_{1}}^{j_{1}}%
,\rho_{-q_{1}}^{i_{1}}\right)  $, $c_{3}(1^{i_{1}},2^{i_{2}},3^{i_{3}})$ for
$c_{3}(\rho_{q_{1}}^{i_{1}},\rho_{q_{2}}^{i_{2}},\rho_{q_{3}}^{i_{3}})$, and
$c_{4}(1^{j_{1}},-1^{i_{1}},2^{j_{2}},-2^{i_{2}})$ for $c_{4}(\rho_{q_{1}%
}^{j_{1}},\rho_{-q_{1}}^{i_{1}},\rho_{q_{2}}^{j_{2}},\rho_{-q_{2}}^{i_{2}})$.\,\,Accordingly,
\[
c_2\left(1^{j1},-1^{i_1}\right)=\overline{c}_2\left(1^{j1},-1^{i_1}\right)
-\overline{c}^0_2\left(1^{j1},-1^{i_1}\right)\!,
\]
\[
c_3(1^{i_1},2^{i_2},3^{i_3})=\overline{c}_3(1^{i_1},2^{i_2},3^{i_3})
-\overline{c}^0_3(1^{i_1},2^{i_2},3^{i_3}),
\]
\[
c_4(1^{j_1},-1^{i_1},2^{j_2},-2^{i_2})=\overline{c}_4(1^{j_1},-1^{i_1},2^{j_2},-2^{i_2})\,-
\]
\[
-\,\overline{c}^0_4(1^{j_1},-1^{i_1},2^{j_2},-2^{i_2}).
\]
Here, the quantities $\overline{c}_{0}^{0}$, $\overline{c}_{2}^{0}$,
$\overline{c}_{3}^{0}$, and $\overline{c}_{4}^{0}$ mean
$\overline{c}_{0}$, $\overline{c}_{2}$, $\overline{c}_{3}$, and
$\overline{c}_{4}$, respectively, in which
$\alpha_{q_{1}}=\alpha_{q_{2}}=\alpha_{q_{3}}=1$.\,\,The quantity
$\overline{c}_{4}(1^{j_{1}},-1^{i_{1}},2^{j_{2}},-2^{i_{2}})$
consists of two terms,
\[
\overline{c}_4(1^{j_1},-1^{i_1},2^{j_2},-2^{i_2})=\overline{c}_4^{(1)}(1^{j_1},-1^{i_1},2^{j_2},-2^{i_2})\,+
\]
\[
+\,\overline{c}_4^{(2)}(1^{j_1},-1^{i_1},2^{j_2},-2^{i_2}),
\]
where
\[
\overline{c}_4^{(1)}(1^{j_1},-1^{i_1},2^{j_2},-2^{i_2})=
\]
\[
=-\frac{1}{64}\frac{\hbar^2}{2m}\frac{(-1)^{i_1+i_2+j_1+j_2}(q_1^2+q_2^2)}{\sh^2(\beta
E_{q_1})\sh^2(\beta E_{q_2})}\,\times
\]
\[
\times\!\sum_{\pm_1}\!\sum_{\pm_2}\!\sum_{\pm_3}\!\pm_1\!\pm_2\!\pm_3\!
\frac{\sh\!\left[\frac{\beta}{2}(E_{q_1}\!\pm_1\!E_{q_1}\!\pm_2\!E_{q_2}\!\pm_3\!E_{q_2})\right]}
{E_{q_1}\pm_1E_{q_1}\pm_2E_{q_2}\pm_3E_{q_2}}\,\times
\]
\[
\times\,
\ch\biggl[\frac{\beta}{2}\biggl(\!(-1)^{i_1}E_{q_1}\pm_1(-1)^{j_1}E_{q_1}\pm_2
\]
\[
\pm_2(-1)^{i_2}E_{q_2} \pm_3(-1)^{j_2}E_{q_2}\!\biggr)\!\biggr]\!.
\]
\[
\overline{c}_4^{(2)}(1^{j_1},-1^{i_1},2^{j_2},-2^{i_2})=
\frac{(-1)^{i_1+i_2+j_1+j_2}}{128\alpha_{q_3}}\left(\!\frac{\hbar^2}{2m}\!\right)^{\!\!2}\!\!\times
\]
\[
\times\,\frac{Q(\tilde{\alpha}_{q_1}, \tilde{\alpha}_{q_2},
\alpha_{q_3})}{\sh^2(\beta E_{q_1}\!)\sh^2(\beta E_{q_2}\!)\sh(\beta
E_{q_3}\!)}\sh\!\left[\frac{\beta}{2}(\tilde{E}_{q_1}\!+\!\tilde{E}_{q_2}\!+\!E_{q_3})\right]\times
\]
\[\times\biggl\{\!\frac{\beta}{\tilde{E}}
\sh\left[\frac{\beta}{2}\left(\!(-1)^{i_1}\tilde{E}_{q_1}+(-1)^{i_2}\tilde{E}_{q_2}-E_{q_3}\!\right)\!\right]
\times
\]
\[
\times\, Q(\tilde{\alpha}_{q_1}, \tilde{\alpha}_{q_2},
\alpha_{q_3})-\frac{2\sh\left[\frac{\beta}{2}\tilde{E}\right]}{\tilde{E}^2}
Q(\tilde{\alpha}_{q_1},\tilde{\alpha}_{q_2}, \alpha_{q_3})\,\times
\]
\[
\times\,\ch\left[\beta\left(\!(i_1-1)\tilde{E}_{q_1}+(i_2-1)\tilde{E}_{q_2}\!\right)\!\right]-\]
\[-\frac{2\sh\left[\frac{\beta}{2}\tilde{E}_{q_2}\right]}{\tilde{E}\tilde{E}_{q_2}}
Q(\tilde{\alpha}_{q_1},-\tilde{\alpha}_{q_2}, \alpha_{q_3})\,\times
\]
\[
\times\,\ch\left[\frac{\beta}{2}\left(\!(-1)^{i_1}\tilde{E}_{q_1}+2(i_2-1)\tilde{E}_{q_2}-E_{q_3}\!\right)\!\right]+\]
\[+\frac{2\sh\left[\frac{\beta}{2}(\tilde{E}_{q_1}+E_{q_3})\right]}{\tilde{E}(\tilde{E}_{q_1}+E_{q_3})}
Q(\tilde{\alpha}_{q_1},-\tilde{\alpha}_{q_2}, \alpha_{q_3})\,\times
\]
\[
\times
\ch\left[\frac{\beta}{2}\left(\!2(i_1-1)\tilde{E}_{q_1}+(-1)^{i_2}\tilde{E}_{q_2}\!\right)\!\right]-
\]
\[
-\frac{2\sh\left[\frac{\beta}{2}\tilde{E}_{q_1}\right]}{\tilde{E}\tilde{E}_{q_1}}
Q(-\tilde{\alpha}_{q_1},\tilde{\alpha}_{q_2}, \alpha_{q_3})\times
\]
\[\times
\ch\left[\frac{\beta}{2}\left(\!(-1)^{i_2}\tilde{E}_{q_2}+2(i_1-1)\tilde{E}_{q_1}-E_{q_3}\!\right)\!\right]+
\]
\[
+\frac{2\sh\left[\frac{\beta}{2}(\tilde{E}_{q_2}+E_{q_3})\right]}{\tilde{E}(\tilde{E}_{q_2}+E_{q_3})}
Q(-\tilde{\alpha}_{q_1},\tilde{\alpha}_{q_2}, \alpha_{q_3})\times
\]
\[
\times
\ch\left[\frac{\beta}{2}\left(\!2(i_2-1)\tilde{E}_{q_2}+(-1)^{i_1}\tilde{E}_{q_1}\!\right)\!\right]-
\]
\[-\frac{2\sh\left[\frac{\beta}{2}E_{q_3}\right]}{\tilde{E}E_{q_3}}
Q(-\tilde{\alpha}_{q_1},-\tilde{\alpha}_{q_2}, \alpha_{q_3})\times
\]
\[
\times
\ch\left[\frac{\beta}{2}\left(\!(-1)^{i_2}\tilde{E}_{q_2}+(-1)^{i_1}\tilde{E}_{q_1}\!\right)\!\right]+\]
\[+\frac{2\sh\left[\frac{\beta}{2}(\tilde{E}_{q_1}+E_{q_2})\right]}{\tilde{E}(\tilde{E}_{q_1}+E_{q_2})}
Q(-\tilde{\alpha}_{q_1},-\tilde{\alpha}_{q_2}, \alpha_{q_3})\times
\]
\[
\times
\ch\left[\frac{\beta}{2}\left(\!2(i_2-1)\tilde{E}_{q_2}+2(i_1-1)\tilde{E}_{q_1}-E_{q_3}\!\right)\!\right]\!
\biggr\}\!.
\]

The coefficients $\overline{c}_{2}(1^{j_{1}},-1^{i_{1}})$ and $\overline
{c}_{0}$ can be expressed in terms of $\overline{c}_{4}^{(1)}(1^{j_{1}%
},-1^{i_{1}},2^{j_{2}},-2^{i_{2}})$ and $\overline{c}_{4}^{(2)}(1^{j_{1}%
},-1^{i_{1}},2^{j_{2}},-2^{i_{2}})$ as follows:%
\[
\overline{c}_2(1^{j_1},-1^{i_1})=2\alpha_{q_2}\sh\left[\beta
E_{q_2}\right]\times
\]
\[
\times\left(\!2\overline{c}_4^{(1)}(1^{j_1},-1^{i_1},2,-2')
+\overline{c}_4^{(2)}(1^{j_1},-1^{i_1},2,-2')\!\right)\!,
\]
\[
\overline{c}_0=4\alpha_{q_1}\alpha_{q_2}\sh\left[\beta
E_{q_1}\right]\sh\left[\beta E_{q_2}\right]\times
\]
\[
\times\left(\!2\overline{c}_4^{(1)}(1,-1',2,-2')+\frac{\overline{c}_4^{(2)}(1,-1',2,-2')}{3}\!\right)\!\!.\]

The quantity $\overline{c}_{3}(1^{i_{1}},2^{i_{2}},3^{i_{3}})$ looks like
\[
\overline{c}_3(1^{i_1},2^{i_2},3^{i_3})\!=\!-\frac{1}{16}\frac{\hbar^2}{2m}\frac{(-1)^{i_1+i_2}(\bf{q}_1\bf{q}_2)}{\sh[\beta
E_{q_1}]\sh[\beta E_{q_2}]\sh[\beta E_{q_3}]}\,\times
\]
\[\times\sum_{\pm_1}\sum_{\pm_2}\left(\alpha_{q_1}\alpha_{q_2}\pm_1\pm_21\right)
\frac{\sh\left[\frac{\beta}{2}\left(\tilde{E}_{q_1}+\tilde{E}_{q_2}+E_{q_3}\right)\right]}{\tilde{E}_{q_1}+\tilde{E}_{q_2}+E_{q_3}}\,\times
\]
\[\times\,
\sh\left[\frac{\beta}{2}\left(\!(-1)^{i_1}\tilde{E}_{q_1}+(-1)^{i_2}\tilde{E}_{q_2}+(-1)^{i_3}E_{q_3}\!\right)\!\right]\!.\]
In the expressions written above, the following notations were introduced:%
\[
\tilde{E}_{q_1}=\pm_1E_{q_1};\quad \tilde{E}_{q_2}=\pm_2E_{q_1};
\]
\[
\tilde{\alpha}_{q_1}=\pm_1\alpha_{q_1};\quad
\tilde{\alpha}_{q_2}=\pm_1\alpha_{q_2};
\]
\[\tilde{E}=\pm_1E_{q_1}\pm_2E_{q_2}+E_{q_3};
\]
\[Q(\tilde{\alpha}_{q_1},\tilde{\alpha}_{q_2},\alpha_{q_3})=(\pm_1\pm_2\alpha_{q_1}\alpha_{q_2}+1)({\bf
q}_1{\bf q}_2)\,+
\]
\[
+\,(\pm_1\alpha_{q_1}\alpha_{q_3}+1)({\bf q}_1{\bf
q}_3)+(\pm_2\alpha_{q_2}\alpha_{q_3}+1)({\bf q}_2{\bf q}_3).
\]

}

\subsubsection*{APPENDIX 2}

{\footnotesize\[
C_2({\mathbf{q}_1})=\overline{C}_2({\mathbf{q}_1})-\overline{C}_2^0({\mathbf{q}_1}),
\]
\[
C_3({\mathbf{q}}_1,{\mathbf{q}}_2,
{\mathbf{q}}_3)=\overline{C}_3({\mathbf{q}}_1,{\mathbf{q}}_2,
{\mathbf{q}}_3)-\overline{C}_3^0({\mathbf{q}}_1,{\mathbf{q}}_2,
{\mathbf{q}}_3),
\]
\[
C_4({\mathbf{q}}_1,{\mathbf{q}}_2)=\overline{C}_4({\mathbf{q}}_1,{\mathbf{q}}_2)
-\overline{C}_4^0({\mathbf{q}}_1,{\mathbf{q}}_2),
\]
where%
\[
 \overline{C}_2({\mathbf{q}_1})=\frac{1}{2}\!\!\!\sum_{i_1=0}^1
\sum_{j_1=0}^1\!\! \overline{c}_2(1^{j1},-1^{i_1});
\]
\[
\overline{C}_3({\mathbf{q}}_1,{\mathbf{q}}_2,
{\mathbf{q}}_3)\!\!=\frac{1}{2}\!\!\sum_{i_1,i_2,i_3=0}^1\!\!
\overline{c}_3(1^{i_1},2^{i_2},3^{i_3});
\]
\[
\overline{C}_4({\mathbf{q}}_1,{\mathbf{q}}_2)\!\!=\frac{1}{2}\!\!\sum_{i_1,i_2=0}^1
\sum_{j_1,j_2=0}^1\!\!\!
\overline{c}_4(1^{j_1},-1^{i_1},2^{j_2},-2^{i_2}).
\]

The notations $\overline{C}_{2}^{0}({\mathbf{q}_{1}})$, $\overline{C}_{3}%
^{0}({\mathbf{q}}_{1},{\mathbf{q}}_{2},{\mathbf{q}}_{3})$, and $\overline
{C}_{4}^{0}({\mathbf{q}}_{1},{\mathbf{q}}_{2})$ mean the quantities
$\overline{C}_{2}({\mathbf{q}_{1}})$, $\overline{C}_{3}({\mathbf{q}}%
_{1},{\mathbf{q}}_{2},{\mathbf{q}}_{3})$, and $\overline{C}_{4}({\mathbf{q}%
}_{1},{\mathbf{q}}_{2})$, respectively, in which the Bogolyubov
factor is equal to unity:
$\alpha_{q_{1}}=\alpha_{q_{2}}=\alpha_{q_{3}}=1$.\,\,The
quantities $\overline{C}_{2}({\mathbf{q}_{1}})$, $\overline{C}_{3}%
({\mathbf{q}}_{1},{\mathbf{q}}_{2},{\mathbf{q}}_{3})$, and $\overline{C}%
_{4}({\mathbf{q}}_{1},{\mathbf{q}}_{2})$ themselves look like%
\[
\overline{C}_2({\bf q_1})=-\frac{1}{4}\sum_{{\bf
q_2}\neq0}\frac{\hbar^2}{2m}\frac{q_1^2+q_2^2}{\alpha_{q_2}\ch^2\left[\frac{\beta}{2}
E_{q_1}\right]\sh[\beta E_{q_2}]}\,\times
\]
\[
\times\biggl\{\!\frac{\beta}{4}\ch[\beta E_{q_2}]-\frac{\sh[\beta
E_{q_2}]}{4E_{q_2}}+\frac{\sh[\beta E_{q_1}]\ch[\beta
E_{q_2}]}{4E_{q_1}}\,-
\]
\[
-\,\frac{\sh[\beta(E_{q_1}+E_{q_2})]}{8(E_{q_1}+E_{q_2})}-\frac{\sh[\beta(E_{q_1}-E_{q_2})]}{8(E_{q_1}-E_{q_2})}\biggr\}+
\frac{1}{16}\left(\!\frac{\hbar^2}{2m}\!\right)^{\!\!2}\!\!\times
\]
\[
\times\!\!\!\mathop{\sum_{\mathbf{q}_2\neq0}\sum_{\mathbf{q}_3\neq0}}\limits_{\mathbf{q}_1+\mathbf{q}_2+\mathbf{q}_3=0}
\frac{Q(\tilde{\alpha}_{q_1},\tilde{\alpha}_{q_2},
\alpha_{q_3})\ch\left[\frac{\beta}{2}(\tilde{E}_{q_2}+E_{q_3})\right]}
{\alpha_{q_2}\alpha_{q_3}\tilde{E}\ch^2\left[\frac{\beta}{2}E_{q_1}\right]\sh\left[\beta
E_{q_2}\right]\sh\left[\beta E_{q_3}\right]}\times
\]
\[
\times\Bigg\{\!\frac{\beta}{4}\sh\left[\frac{\beta}{2}(\tilde{E}_{q_2}+E_{q_3})\right]Q(\tilde{\alpha}_{q_1},\tilde{\alpha}_{q_2},
\alpha_{q_3})\,+
\]
\[+\,\frac{1}{2}\sh\left[\frac{\beta}{2}\tilde{E}\right]\sh\left[\frac{\beta}{2}\tilde{E}_{q_1}\right]
\times
\]
\[\times\Bigg(\!\frac{Q(-\tilde{\alpha}_{q_1},\tilde{\alpha}_{q_2},
\alpha_{q_3})}{\tilde{E}}+\frac{Q(\tilde{\alpha}_{q_1},\tilde{\alpha}_{q_2},
\alpha_{q_3})}{\tilde{E}_{q_1}}\!\Bigg)-
\]
\[
-\,\frac{1}{2}\sh\left[\frac{\beta}{2}(\tilde{E}_{q_1}+\tilde{E}_{q_2})\right]
\sh\left[\frac{\beta}{2}(\tilde{E}_{q_1}+E_{q_3})\right]\times
\]
\[
\times\Bigg(\!\frac{Q(-\tilde{\alpha}_{q_1},-\tilde{\alpha}_{q_2},
\alpha_{q_3})}{\tilde{E}_{q_1}+\tilde{E}_{q_2}}+\frac{Q(\tilde{\alpha}_{q_1},\tilde{\alpha}_{q_2},
\alpha_{q_3})}{\tilde{E}_{q_1}+E_{q_3}}\!\Bigg)-
\]
\[
-\,\frac{1}{2}\sh\left[\frac{\beta}{2}\tilde{E}_{q_2}\right]
\sh\left[\frac{\beta}{2}E_{q_3}\right]\times
\]
\[
\times \Bigg(\!\frac{Q(\tilde{\alpha}_{q_1},-\tilde{\alpha}_{q_2},
\alpha_{q_3})}{\tilde{E}_{q_2}}+\frac{Q(-\tilde{\alpha}_{q_1},-\tilde{\alpha}_{q_2},
\alpha_{q_3})}{E_{q_3}}\!\Bigg)\!\!\Bigg\}\!.
\]
%5
\[
\overline{C}_3({\bf q_1},{\bf q_2},{\bf
q_3})=-\frac{1}{48}\frac{\hbar^2}{2m}
\frac{\sh\left[\frac{\beta}{2}\tilde{E}\right]}{\tilde{E}\prod\limits_{j=1}^3\ch\left[\frac{\beta}{2}E_{q_j}
\right]}\,\times
\]
\begin{equation}
\times\, Q(\tilde{\alpha}_{q_1},\tilde{\alpha}_{q_2}, \alpha_{q_3}).
\end{equation}
\[
\overline{C}_4({\bf q_1},{\bf q_2})=-\frac{1}{16}\sum_{{\bf
q_1}\neq0}\sum_{{\bf
q_2}\neq0}\frac{\hbar^2}{2m}\frac{(q_1^2+q_2^2)}{\ch^2\left[\frac{\beta}{2}
E_{q_1}\right]}\,\times
\]
\[\times\,\frac{1}{\ch^2\left[\frac{\beta}{2}
E_{q_2}\right]}\Bigg\{\!\frac{\beta}{4}+\frac{\sh[\beta
E_{q_2}]}{4E_{q_2}}+\frac{\sh[\beta E_{q_1}]\ch[\beta
E_{q_2}]}{4E_{q_1}}\,+
\]
\[
+\,\frac{\sh[\beta(E_{q_1}+E_{q_2})]}{8(E_{q_1}+E_{q_2})}\!+\!\frac{\sh[\beta(E_{q_1}-E_{q_2})]}{8(E_{q_1}-E_{q_2})}\biggr\}
+\frac{1}{64}\left(\frac{\hbar^2}{2m}\right)^{\!\!2}\!\times
\]
\[
\times
\mathop{\sum_{\mathbf{q}_3\neq0}}\limits_{\mathbf{q}_1+\mathbf{q}_2+\mathbf{q}_3=0}
\!\!\!\frac{Q(\tilde{\alpha}_{q_1},\tilde{\alpha}_{q_2},
\alpha_{q_3})}
{\alpha_{q_3}\tilde{E}\ch^2\left[\frac{\beta}{2}E_{q_1}\right]\ch^2\left[\frac{\beta}{2}
E_{q_2}\right]\ch\left[\frac{\beta}{2} E_{q_3}\right]}\times
\]
\[
\times\Bigg\{\!\!\Bigg(\!\frac{\beta}{4}\!\ch\left[\frac{\beta}{2}E_{q_3}\right]
\!-\!\frac{\sh\left[\frac{\beta}{2}\tilde{E}\right]}{2\tilde{E}}
\ch\left[\frac{\beta}{2}(\tilde{E}_{q_1}\!+\!\tilde{E}_{q_2})\right]\!\!\Bigg)
\times
\]
\[
\times\, Q(\!\tilde{\alpha}_{q_1}\!,\tilde{\alpha}_{q_2}\!,
\alpha_{q_3}\!)\!+\!\Bigg(\!\frac{\sh\!\left[\frac{\beta}{2}\tilde{E}_{q_2}\right]}{2\tilde{E}_{q_2}}
\ch\!\left[\frac{\beta}{2}(\tilde{E}_{q_2}\!+\!E_{q_3})\right]-
\]
\[-\,
\frac{\sh\!\left[\frac{\beta}{2}(\tilde{E}_{q_1}\!+\!E_{q_3})\right]}{2(\tilde{E}_{q_1}+E_{q_3})}
\ch\!\left[\frac{\beta}{2}\tilde{E}_{q_2}\right]\!\!\Bigg)\!
Q(\tilde{\alpha}_{q_1},-\tilde{\alpha}_{q_2}, \alpha_{q_3})\,+
\]
\[
+\Bigg(\!\frac{\sh\left[\frac{\beta}{2}\tilde{E}_{q_1}\right]}{2\tilde{E}_{q_1}}
\ch\left[\frac{\beta}{2}(\tilde{E}_{q_1}+E_{q_3})\right]-
\]
\[
-\frac{\sh\!\left[\frac{\beta}{2}(\tilde{E}_{q_2}\!+\!E_{q_3})\right]}{2(\tilde{E}_{q_2}+E_{q_3})}
\ch\!\left[\frac{\beta}{2}\tilde{E}_{q_1}\right]\!\!\Bigg)\!Q(-\tilde{\alpha}_{q_1},\tilde{\alpha}_{q_2},
\alpha_{q_3})\,+
\]
\[
+\Bigg(\!-\frac{\sh\!\left[\frac{\beta}{2}E_{q_3}\right]}{2E_{q_3}}
\!+\!\frac{\sh\!\left[\frac{\beta}{2}(\tilde{E}_{q_1}\!+\!\tilde{E}_{q_2})\right]}{2(\tilde{E}_{q_1}+\tilde{E}_{q_2})}
\ch\!\left[\frac{\beta}{2}\tilde{E}\right]\!\!\Bigg)\times
\]
\[
\times\, Q(-\tilde{\alpha}_{q_1},-\tilde{\alpha}_{q_2},
\alpha_{q_3})\!\Bigg\}\!.
\]

}

\vspace*{-5mm}
\rezume{%
І.О.\,Вакарчук, О.І.\,Григорчак}{СТРУКТУРНІ ФУНКЦІЇ\\ БАГАТОБОЗОННОЇ
СИСТЕМИ\\ ІЗ ВРАХУВАННЯМ ПРЯМИХ ТРИ-\\ ТА ЧОТИРИЧАСТИНКОВИХ
КОРЕЛЯЦІЙ} {На основі виразу для матриці густини взаємодіючих
бозе-частинок в координатному зображенні  із врахуванням прямих три-
і чотиричастинкових  кореляцій [І. О. Вакарчук, О. І. Григорчак,
Журн. фіз. досл. {\bf 3}, 3005 (2009)] були розраховані \mbox{дво-,}
три- і чотиричастинкові структурні фактори рідкого $^4$He в
наближенні ``однієї суми за хвильовим вектором'' для широкого
інтервалу температур. В границі низьких температур отриманий вираз
для двочастинкового структурного фактора переходить в уже відомий. В
границі високих температур вирази для дво-, три- і чотиричастинкових
структурних факторів редукуються до структурних факторів ідеального
бозе-газу. Результати роботи можуть бути застосовані для  розрахунку
термодинамічних функцій рідкого $^4$He і знаходження температурної
залежності швидкості першого звуку в багатобозонній системі.}


\begin{thebibliography}{99}                                                                                               %


\bibitem {BPS}I.V. Bogoyavlenskii, A.V. Puchkov, and A.~Skomorokhov, Physica~B
\textbf{284--288}, 25 (2000).

\bibitem {FeFam}Fereydoon Family, Physica B\,+\,C \textbf{107}, 699 (1981).

\bibitem {ZPSS}F. Zambelli, L. Pitaevskii, D.M.~Stamper-Kurn, and
S.~Stringari, Phys. Rev.~A \textbf{61}, 063608 (2000).

\bibitem {KrLi}E. Krotscheck and T. Lichtenegger, J.~Low Temp. Phys.
\textbf{178}, 61 (2015).

\bibitem {HHKP}R. Hobbiger, R. Holler, E.~Krotscheck, and M.~Panholzer, J.~Low
Temp. Phys. \textbf{169}, 350 (2012).

\bibitem {SPA}V. Sorkin, E. Polturak, and J.~Adler, J.~Low Temp. Phys.
\textbf{143}, 141 (2006).

\bibitem {KrTy}E. Krotscheck and C.J. Tymczak, Phys. Rev.~B \textbf{45}, 217 (1992).

\bibitem {MiSc}K,~Miyazaki and I.M. de Schepper, Phys. Rev. E \textbf{63}
060201 (2001).

\bibitem {BTV}V.B. Bobrov, S.A. Trigger, and Yu.P. Vlasov, Phys. B: Cond. Matt.
\textbf{203}, 95 (1994).

\bibitem {DBRF}J. Dawidowski, F. J. Bermejo, M. L. Ristig, B. F\.{a}k, C.~Cab\-rillo, R. Fern\'{a}ndez-Perea, K. Kinugawa, and J.~Cam\-po, Phys. Rev. B
\textbf{69}, 014207 (2004).

\bibitem {HaJa}F. Hayot and C. Jayaprakash, Phys. Rev. E \textbf{57}, 4867(R) (1998).

\bibitem {Svensson}E.C. Svensson, V.F. Sears, A.D.B. Woods, and P.~Martel,
Phys. Rev. B \textbf{21}, 3638 (1980).\vspace*{0.4mm}

\bibitem {Robkoff}H.N. Robkoff and R.B. Hallock, Phys. Rev. B \textbf{24}, 159 (1981).\vspace*{0.4mm}

\bibitem {CBA}F.~Caupin, J.~Boronat, and K.H. Andersen, J.~Low Temp. Phys.
\textbf{152}, 108 (2008).\vspace*{0.4mm}

\bibitem {SOKD}J. Steinhauer, R. Ozeri, N. Katz, and N. Davidson, Phys. Rev. A
\textbf{72}, 023608 (2005).\vspace*{0.4mm}

\bibitem {ChCo}Chia-Wei Woo and R.L. Coldwell, Phys. Rev. Lett. \textbf{29},
1062 (1972).\vspace*{0.4mm}

\bibitem {ALM}N.G. Almarza, E. Lomba, and D. Molina, Phys. Rev. E \textbf{70},
021203 (2004).\vspace*{0.4mm}

\bibitem {ZTG}P.~Zi\'{n}, M.~Trippenbach, and M.~Gajda, Phys. Rev. A
\textbf{69}, 023614 (2004).\vspace*{0.4mm}

\bibitem {VakUhn79-80}I.A.~Vakarchuk and I.R. Yukhnovskii, Theor. Math. Phys.
\textbf{40}, 626 (1979); \textbf{42}, 73 (1980).\vspace*{0.4mm}

\bibitem {VHU79}I.A.~Vakarchuk, A.L. Gonopolskii, and I.R.~Yukhnovskii, Theor. Math. Phys. \textbf{41}, 896 (1979).\vspace*{0.4mm}

\bibitem {Vak85-90}I.A.~Vakarchuk, Theor. Math. Phys. \textbf{65}, 1164 (1985);
\textbf{82}, 308 (1990).\vspace*{0.4mm}

\bibitem {VH1}I.A.~Vakarchuk and P.A. Glushak, Theor. Math. Phys. \textbf{75},
399 (1988).\vspace*{0.4mm}

\bibitem {Hlushak}P.A. Glushak, \textit{Research of Equilibrium Properties of
Superfluid Helium-4 at Low Temperatures,} Ph.D. thesis (Lviv, 1992)
(in Russian).\vspace*{0.4mm}

\bibitem {VPR2007}I.O.~Vakarchuk, R.O.~Prytula, and A.A.~Rovenchak, J.~Phys. Stud. \textbf{11}, 259 (2007).\vspace*{0.4mm}

\bibitem {VP2008-09}I.O.~Vakarchuk and R.O.~Prytula, J. Phys. Stud.
\textbf{12}, 4001 (2008); \textbf{13}, 2003 (2009).\vspace*{0.4mm}

\bibitem {exps1}F.K. Achter and L. Meyer, Phys. Rev. \textbf{188}, 291 (1969).\vspace*{0.4mm}

\bibitem {Temperly}\textit{Physics of Simple Liquids,} edited by
H.N.V.~Temperley, J.S.~Rowlinson, and G.S.~Rushbrooke
(North-Holland, Amsterdam, 1968).\vspace*{0.4mm}

\bibitem {Krokston}C.A.~Croxton, \textit{Liquid State Physics: A Statistical
Mechanical Introduction} (Cambridge Univ. Press, Cambridge,
2009).\vspace*{0.5mm}

\bibitem {VakHryh1}I.O.~Vakarchuk and {O.I. Hryhorchak}, J. Phys. Stud.
\textbf{3}, 3005 (2009).

\bibitem {VakHryh2}I.O.~Vakarchuk and {O.I. Hryhorchak}, Visn. L'viv. Univ.
Ser. Fiz. \textbf{46}, 3 (2011).

\bibitem {VBR}I.O. Vakarchuk, V.V. Babin, and A.A.~Rovenchak, J. Phys. Stud.
\textbf{4}, 16 (2000).

\bibitem {Rov-dys}A.A.~Rovenchak, \textit{Self-Consistent Calculation of
Interato\-mic Potentials and Thermodynamic Functions of Helium-4 in
Superfluid and Normal Phases,} Ph.D. thesis (Lviv, 2003) (in
Ukrainian).

\bibitem {HP2014}I.O.~Vakarchuk, {O.I. Hryhorchak}, V.S.~Pastukhov, and
R.O.~Prytula, arXiv:1506.03317 (2015).

\bibitem {Vak2004}I.O. Vakarchuk, J. Phys. Stud. \textbf{8}, 223 (2004).

\bibitem {Vstup}I.O.~Vakarchuk, \textit{Introduction to Many-Body Problem}
(Lviv National Univ., Lviv, 1999) (in Russian).

\bibitem {DonBar}R.J. Donnelly and C.F. Barenghi, J. Phys. Chem. Ref. Data
\textbf{27}, 1217 (1998).\vspace*{-2mm}
\begin{flushright}
{\footnotesize Received 01.04.15.\\ Translated from Ukrainian by
O.I.~Voitenko}
\end{flushright}
\end{thebibliography}
\end{document}